\documentclass[reprint,aps,onecolumn,11pt,notitlepage]{revtex4-1}
\usepackage[english]{babel}
\usepackage{amsmath,amssymb,graphicx,bm}
\usepackage[colorlinks=true,citecolor=blue,linkcolor=blue,urlcolor=blue]{hyperref} 
\usepackage[justification=raggedright,font=small,labelfont=bf]{caption}
\usepackage{setspace}
 
\begin{document}
\onehalfspacing

\title{Non-linear corrections of overlap reduction functions for pulsar timing arrays}
 
\author{Qing-Hua Zhu}
\email{zhuqh@itp.ac.cn} 
\affiliation{CAS Key Laboratory of Theoretical Physics, 
		Institute of Theoretical Physics, Chinese Academy of Sciences,
		Beijing 100190, China}
\affiliation{School of Physical Sciences,   
		University of Chinese Academy of Sciences, 
		No. 19A Yuquan Road, Beijing 100049, China}
	

\date{\today}
	
\begin{abstract}
Recent pulsar timing array experiments have reported a hint of gravitational stochastic background in nHz band frequency. Further confirmation might rely on whether the signature of the signals is consistent with that sketched by overlap reduction functions, known as Hellings-Downs curves. This paper investigates the non-linear corrections of overlap reduction functions in the presence of non-Gaussianity, in which the self-interaction of gravity is first taken into account. Expanding Einstein field equations and geodesic equations into the non-linear regime, we obtain non-linear corrections for the timing residuals of pulsar timing, and theoretically study the overlap reduction functions for pulsar timing arrays. In the event of considerable correction from the three-point correlations of gravitational waves, the shapes of the overlap reduction functions with non-linear corrections can be differentiated from Hellings-Downs curves.
\end{abstract}

\maketitle 

\section{Introduction}


The stochastic gravitational wave background (SGWB) might contain lots of information about the Universe. It can be originated from inflationary GWs \cite{,Grishchuk:1974ny,Starobinsky:1979ty,Caprini:2018mtu,Vagnozzi:2020gtf,Benetti:2021uea}, produced from early-time phase transitions \cite{Witten:1984rs,Hogan:1986qda,NANOGrav:2021flc}, sourced by cosmic string \cite{Vilenkin:1981bx,Hogan:1984is,Vachaspati:1984gt,LIGOScientific:2021nrg}, or formed by superpositions of unresolved individual GW sources such as binary systems \cite{Schneider:2000sg,Farmer:2003pa,Sesana:2008mz,LIGOScientific:2016fpe,LIGOScientific:2017zlf}, core-collapse supernovae \cite{10.1093/mnras/283.2.648,Ferrari:1998ut,Buonanno:2004tp,Finkel:2021zgf}, and deformed rotating neutron stars \cite{Owen:1998xg,Ferrari:1998ut}. 
In the $10\rm Hz$--$1\rm kHz$ frequency band, the ground-based GW detectors LIGO/Virgo/KAGRA have established an upper limit of SGWBs \cite{KAGRA:2021kbb,KAGRA:2021mth}.
In $\rm nHz$ frequency band, the international timing pulsar array projects (IPTA) \cite{Jenet:2009hk,Hobbs:2013aka,2008AIPC..983..633J} found and confirmed a common spectrum process from the pulsar-timing data sets, and suggested that further evidence for SGWBs might rely on its angular correlation signature \cite{NANOGrav:2020bcs,Chen:2021rqp,Goncharov:2021oub}.

The angular correlations of the output of a pair of GW detectors indicate characteristic signature of GWs, which is formulated by overlap reduction functions (ORFs). For pulsar timing arrays (PTAs), the ORFs of GWs are known as Hellings-Downs curve \cite{Hellings:1983fr}. Inspired by recent observations on the SGWBs \cite{NANOGrav:2020bcs,Chen:2021rqp,Goncharov:2021oub}, there would be of physical interest to investigate the physical causes underlying deviations from the Hellings-Downs curve. Specifically, it could be originated from the SGWBs beyond isotropy approximation \cite{Mingarelli:2013dsa,Himemoto:2019iwd}, polarized SGWBs \cite{Omiya:2021zif,Omiya:2020fvw,Chu:2021krj}, non-tensor modes from modified gravity \cite{Nishizawa:2009bf,Lee:2010cg,Boitier:2020xfx,Liang:2021bct,Chen:2021wdo,Boitier:2021rmb}, or simply a careful calculation on the pulsar terms \cite{Mingarelli:2014xfa,Boitier:2020rzg,Hu:2022ujx}. Recently, there was also study on the non-linear corrections for PTAs from the higher order effect of gravity \cite{Tasinato:2022xyq}. This correction of ORFs was shown to be order-one in the presence of non-Gaussianity of the GWs. 
In this paper, we will extend the study to the non-linear corrections of the ORFs, in which the self-interaction of gravity is taken into account. 

Owing to the self-interaction of gravity, linear-order GWs have the capacity to generate non-linear GWs even in vacuum. It can influence pulsar timing and consequently alter the response of GW detectors.
To obtain a solid derivation for GW detectors to the non-linear regime, we utilize perturbed Einstein field equations for the evolution of metric perturbations, and perturbed geodesic equations to calculate timing residuals of pulsar timing. 
We compute the ORFs with the non-linear corrections, and study the shapes of non-linear ORFs with different phenomenological parameters of non-Gaussianity. 
 
The rest of the paper is organized as follows. In Sec.~\ref{II}, the evolutions of metric perturbations to the second order are presented. In Sec.~\ref{III}, based on the propagation of light formulated by the perturbed geodesic equations to the second order, we obtain timing residuals of pulsar timing with non-linear corrections. In Sec.~\ref{IV}, we compute ORFs with different shapes of non-Gaussianity, and show the deviation from the Hellings-Downs curves. In Sec.~\ref{V}, conclusions and discussions are summarized.

\section{Propagation of secondary gravitational wave within PTAs} \label{II}

From the theory of perturbations in general relativity, the first-order GWs, characterized by the transverse-traceless modes of metric perturbations, have the capacity to produce secondary effects from nonvanishing higher-order metric perturbations. 
Due to the fact that fluctuations in space-time can influence the propagation of light, it is inevitable that there would be corrections to the response of GW detectors in a non-linear regime. In this section, to study the secondary effects of GWs on PTAs, we will show how the non-linear order metric perturbations freely propagate in vacuum.

The perturbed metric in Minkowski background is given by \cite{Weinberg:2008zzc}
\begin{eqnarray}
	{\rm d}s^2 = -{\rm d}t^2+ \left( \delta_{i  j} (1- \psi^{(2)}) + 
	\partial_i \partial_j E^{(2)} + \frac{1}{2}\partial_i C^{(2)}_j + \frac{1}{2}\partial_j C_i^{(2)}
	+ h_{i  j}^{(1)}+\frac{1}{2}h_{i  j}^{(2)}\right){\rm d}x^i{\rm d}x^j~, \label{3a}
\end{eqnarray}
where $\delta_{ij}$ is Kronecker symbol, $h^{(1)}_{ij}$ is the first order transverse-traceless metric perturbation known as GWs, and $h^{(2)}_{ij}$, $C^{(2)}_j$, and $\psi^{(2)}$ are the second order tensor, vector, and scalar perturbations, respectively.
Because GW detectors are considered in the freely-falling frame, we have adopted Synchronous gauge for the perturbed metric in Eq.~(\ref{3a}).

From Einstein field equations for the first order GWs, the motion of equations reduce to wave equations,
\begin{eqnarray}
	{h_{i  j}^{(1)}}'' - \Delta h_{i  j}^{(1)} & = & 0~. \label{1}
\end{eqnarray}
Thus, the solutions can be given by
\begin{eqnarray}
	h_{i  j}^{(1)} & = & \int \frac{{\rm d}^3 k}{(2 \pi)^3} \bar{h}^{(1)}_{i
	j, \bm{k}} e^{- i \left( k  t - \bm{k} \cdot \bm{x}
			\right)}~, \label{2}
\end{eqnarray}
where the $h_{i  j, \bm{k}}^{(1)}$ is the Fourier mode of $\bar{h}^{(1)}_{ij}$, which contains the physical information before the GWs reaching the detectors.

Since the GW detectors have responses to tensor, vector, and scalar modes of the metric perturbations, we should further consider all the second-order metric perturbations. 
Using the metric in Eq.~(\ref{3a}), the perturbed Einstein field equations in the second order take the form of \cite{Chang:2020tji,Zhou:2021vcw}
\begin{subequations}
\begin{eqnarray}
	{h_{i  j}^{(2)}}'' - \Delta h_{i  j}^{(2)} & = & -
	\Lambda_{i  j}^{a  b} \mathcal{S}_{a  b}~,\\
	{C^{(2)}_j}'' & = & -\mathcal{V}_j^{a  b} \mathcal{S}_{a
		b}~,\\
	{- 2 \psi^{(2)}}'' + \Delta \left( {E^{(2)}}'' + \psi^{(2)} \right) & = & -
	S^{(\Psi), a  b} \mathcal{S}_{a  b}~,\\
	{E^{(2)}}'' + \psi^{(2)} & = & - S^{(E), a  b} \mathcal{S}_{a
		b}~,
\end{eqnarray} \label{3}
\end{subequations}
where $\Lambda_{i  j}^{a  b}$, $ \mathcal{V}_j^{a
	b}$, $S^{(\Psi), a  b}$ and $S^{(E), a  b}$ are
helicity decomposition operators \cite{Weinberg:2008zzc}, and the source on the right hand side of Eq.~(\ref{3}) is given by
\begin{eqnarray}
	\mathcal{S}_{a  b} & = & - 2 \delta^{c  d} \partial_0
	h^{(1)}_{a  c} \partial_0 h_{b  d}^{(1)} + 2 h^{(1), c
			d} \partial_c \partial_d h^{(1)}_{a  b} - 2 h^{(1), c
			d} \partial_c \partial_a h^{(1)}_{d  b} \nonumber\\
	&& - 2 h^{(1), c
			d} \partial_c \partial_b h^{(1)}_{d  a} - 2 \partial^d
	h^{(1)}_{a  c} \partial^c h^{(1)}_{b  d} + 2 \delta^{c
		d} \partial_j h_{a  d}^{(1)} \partial^j h_{b
			c}^{(1)} + \partial_a h^{(1), c  d} \partial_b h^{(1)}_{c
			d} + 2 h^{(1), c  d} \partial_a \partial_b h^{(1)}_{c  d} \nonumber\\
	&& +
	\delta_{a  b} \left( \frac{3}{2} \partial_0 h^{(1)}_{c  d}
	\partial_0 h^{(1), c  d} + 2 h_{c  d}^{(1)} \partial_0^2
	h^{(1), c  d} - 2 h_{c  d}^{(1)} \Delta h^{(1), c
				d} + \partial_j h^{(1)}_{c  d} \partial^d h^{(1), c  j} -
	\frac{3}{2} \partial_j h_{c  d}^{(1)} \partial^j h^{(1), c
				d} \right)~. \nonumber\\ \label{4}
\end{eqnarray}
Differing from the linear regime, the second metric perturbations are sourced by the first-order GWs.
By making use of Eq.~(\ref{2}) and (\ref{3}), we obtain the solutions in the form of
\begin{subequations}
\begin{eqnarray}
	\psi^{(2)} \left( t, \bm{x} \right) & = & \int \frac{{\rm d}^3 k}{(2\pi)^3} \left\{ \int \frac{{\rm d}^3 p}{(2 \pi)^3} \left\{ \hat{F}_{\psi}^{ab} \bar{\mathcal{S}}_{a  b} \left( \bm{k}, \bm{p}\right) e^{- i \left( \left| \bm{k} - \bm{p} \right| + p \right) t}
	\right\} e^{i \bm{k} \cdot \bm{x}} \right\}~,\\
	E^{(2)} \left( t, \bm{x} \right) & = & \int \frac{{\rm d}^3 k}{(2 \pi)^3}
	\left\{ \int \frac{{\rm d}^3 p}{(2 \pi)^3} \left\{ \hat{F}_E^{a  b}
	\bar{\mathcal{S}}_{a  b} \left( \bm{k}, \bm{p} \right) e^{-
			i \left( \left| \bm{k} - \bm{p} \right| + p \right) t} \right\} e^{i
			\bm{k} \cdot \bm{x}} \right\}~,\\
	C_j^{(2)} \left( t, \bm{x} \right) & = & \int \frac{{\rm d}^3 k}{(2
		\pi)^3} \left\{ \int \frac{{\rm d}^3 p}{(2 \pi)^3} \left\{ \hat{F}_{C,
		j}^{a  b} \bar{\mathcal{S}}_{a  b} \left(
	\bm{k}, \bm{p} \right) e^{- i \left( \left| \bm{k} - \bm{p}
			\right| + p \right) t} \right\} e^{i \bm{k} \cdot \bm{x}} \right\}~,\\
	h_{i  j}^{(2)} \left( t, \bm{x} \right) & = & \int
	\frac{{\rm d}^3 k}{(2 \pi)^3} \left\{ \int \frac{{\rm d}^3 p}{(2 \pi)^3}
	\left\{ \hat{F}_{h,  i  j}^{a  b}
	\bar{\mathcal{S}}_{a  b} \left( \bm{k}, \bm{p} \right) e^{-
			i \left( \left| \bm{k} - \bm{p} \right| + p \right) t} \right\} e^{i
			\bm{k} \cdot \bm{x}} \right\}~.
\end{eqnarray} \label{5}
\end{subequations}
The decomposition operators in momentum space $\hat{F}_{*}^{a  b}\left( \bm{k}, \bm{p} \right)$ ($*=h,C,\phi,E$), and the source $\bar{\mathcal{S}}_{i  j} \left( \bm{k}, \bm{p} \right)$ in Eqs.~(\ref{5}) are presented as follows,
\begin{subequations}
\begin{eqnarray}
	\hat{F}_{h, i  j}^{a  b}\left( \bm{k}, \bm{p} \right) & \equiv & - \frac{\Lambda^{a
			b}_{i  j} \left( \bm{k} \right)}{k^2 - \left( \left| \bm{k}- \bm{p} \right| + p \right)^2}~,\\
	\hat{F}_{C,  j}^{a  b}\left( \bm{k}, \bm{p} \right) & \equiv & \frac{\mathcal{V}^{a
			b}_j \left( \bm{k} \right)}{\left( \left| \bm{k} - \bm{p}
		\right| + p \right)^2}~,\\
	\hat{F}^{a  b}_{\psi}\left( \bm{k}, \bm{p} \right) & \equiv & - \frac{S^{(\Psi), a  b} \left(
		\bm{k} \right) + k^2 S^{(E), a  b} \left( \bm{k} \right)}{2
		\left( \left| \bm{k} - \bm{p} \right| + p \right)^2}~,\\
	\hat{F}^{a  b}_E\left( \bm{k}, \bm{p} \right) & \equiv & \frac{S^{(E), a  b} \left( \bm{k}
		\right)}{\left( \left| \bm{k} - \bm{p} \right| + p \right)^2} -
	\frac{S^{(\Psi), a  b} \left( \bm{k} \right) + k^2 S^{(E), a
				b} \left( \bm{k} \right)}{2 \left( \left| \bm{k} -
		\bm{p} \right| + p \right)^4}~,
\end{eqnarray}	\label{7}
\end{subequations}
and
\begin{eqnarray}
	\bar{\mathcal{S}}_{i  j} \left( \bm{k}, \bm{p} \right) & = &
	f^{ c  d  a  b}_{i  j} \left(
	\bm{k}, \bm{p} \right) \bar{h}^{(1)}_{c  d, \bm{k} -
	\bm{p}} \bar{h}_{a  b, \bm{p}}^{(1)}~, \label{8}
\end{eqnarray}
where the $f_{i j}^{b c l m} \left( \bm{k},
	\bm{p} \right)$ is defined with
\begin{eqnarray}
	f_{i j}^{b c l m} \left( \bm{k},
	\bm{p} \right) & = & \delta^{b l} \delta^{c m} \left(
	- 3 \delta_{i j} \left( p \left| \bm{k} - \bm{p} \right| -
	p \cdot (k - p) \right) + p_i (k_j - 2 p_j) + k_i (p_j - 2 k_j) \right)\nonumber\\
	& & + 2 \left(p^b (- \delta_i^l \delta_j^m p^c + \delta^{c m}
	(\delta_j^l p_i + \delta_i^l p_j + \delta_{i j} (p^l - k^l)))\right.\nonumber\\
	& & + \delta_j^b \left( \delta^{c m} \left( \delta_i^l \left( p
	\left| \bm{k} - \bm{p} \right| - p \cdot (k - p) \right) + k_i (k^l
	- p^l) + p_i (p^l - k^l)\right) + \delta_i^l p^c (k^m - p^m) \right)\nonumber\\
	& & + \delta^b_i \left( \delta^{c m} \left( \delta_j^l \left( p
	\left| \bm{k} - \bm{p} \right| - p \cdot (k - p) \right) + k_j (k^l
	- p^l) + p_j (p^l - k^l)\right) + \delta_j^l p^c (k^m - p^m) \right)\nonumber\\
	& & \left.- \delta_i^b \delta_j^c (k^l k^m - 2 k^l p^m + p^l p^m)\right)~.
\end{eqnarray} 
As shown in Eqs.~(\ref{5}), the explicit solutions of perturbations $\psi^{(2)}$, $E^{(2)}$, $C_j^{(2)}$ and $h^{(2)}_{i  j}$ can be obtained with the known $h^{(1)}_{i    j}$ in Eq.~(\ref{2}). 

We consider the non-linear corrections for GW detectors originated from the gravity self-interaction in vacuum. In this case, all the second-order perturbations are generated by the first-order one. In fact, there could be second-order perturbations that are independent of the $h^{(1)}_{ij}$ \cite{Baumann:2007zm}. 
Owing to the fact that the response elicited by the second-order perturbations of this type to the GW detector does not differ from the response for first-order GWs, this situation was not taken into consideration in the present study.


\section{Perturbed geodesic equations} \label{III}

The space-time fluctuations can affect the time of arrival of radio beams from a pulsar. Thus, via monitoring pulsar timing, the PTA observations can reflect the space-time fluctuations over the Universe in principle. To formulate it, and extend it to the non-linear regime, we will calculate the perturbed geodesic equations to the second order.

Based on geodesic equations in Minkowski background, namely,
\begin{eqnarray}
	P^{\mu} \partial_{\mu} P^{\nu} & = & 0~,
\end{eqnarray}
one can obtain the 4-momentum of a light ray,
\begin{eqnarray}
	P^{\mu} & = & P^0 (1, - \hat{n}^j)~, \label{11}
\end{eqnarray}
where $\hat{n}^j$ is a constant unit vector, and can be used to locate a pulsar. By making use of the above 4-momentums, one can obtain trajectories of light rays from the location of a pulsar to the detectors. The pulsar emits a radio beam at the event $(t-L,	L \hat{n}^j)$, and the beam is detected on the earth at the event of $(t, 0)$, where the $L$ is the distance between the pulsar and the detectors.

From the first-order perturbed geodesic equations,
\begin{eqnarray}
	0 & = & \delta P^{\mu} \partial_{\mu} P^{\nu} + P^{\mu} \partial_{\mu} \delta
	P^{\nu} + g^{\nu \rho} \left( \partial_{\mu} h_{\lambda \rho}^{(1)} -
	\frac{1}{2} \partial_{\rho} h_{\mu \lambda}^{(1)} \right) P^{\mu} P^{\lambda}~, \label{13a}
\end{eqnarray}
we can evaluate it by using Eq.~(\ref{11}), namely,
\begin{subequations}
\begin{eqnarray}
	(\partial_0 - \hat{n} \cdot \partial) \left( \frac{\delta P^0}{P^0} \right)
	& = & - \frac{1}{2} \hat{n}^a \hat{n}^b \partial_0 h_{a  b}^{(1)}~,\label{14a}\\
	(\partial_0 - \hat{n} \cdot \partial) \left( \frac{\delta P^j}{P^0} \right)
	& = & \delta^{j  b} \hat{n}^a \left( \partial_0 h_{a  b}^{
				(1)} - \hat{n}^c \left( \partial_c h_{a  b}^{(1)} - \frac{1}{2}
		\partial_b h_{ a  c}^{(1)} \right) \right)~,
\end{eqnarray} \label{13}
\end{subequations}
where $\delta P^\mu$ is the first order perturbed 4-momentum.
The temporal and spatial components of the above perturbed geodesic equations take different forms, because the time components of $h_{\mu\nu}^{(1)}$ in Eq.~(\ref{13a}) vanish in the Synchronous gauge. From Eqs.~(\ref{13}), the solutions in Fourier space are obtained, 
\begin{subequations}
	\begin{eqnarray}
	\frac{\delta P^0_{\bm{k}}}{P^0} \equiv \mathcal{K}^{0, a  b}
	\left( \bm{k}, \hat{n} \right) \bar{h}_{a  b, \bm{k}}^{(1)}
	e^{- i  k  t} & = & - \frac{1}{2 (1 + \hat{n} \cdot
		\hat{k})} \hat{n}^a \hat{n}^b \bar{h}_{a  b, \bm{k}}^{(1)} e^{-
			i  k  t}~,\\
	\frac{\delta P^j_{\bm{k}}}{P^0} \equiv \mathcal{K}^{j, a  b}
	\left( \bm{k}, \hat{n} \right) \bar{h}_{a  b, \bm{k}}^{(1)}
	e^{- i  k  t} & = & \left( \delta^{j  b} \hat{n}^a -
	\frac{\hat{k}^j \hat{n}^a \hat{n}^b}{2 (1 + \hat{n} \cdot \hat{k})} \right)
	\bar{h}_{a  b, \bm{k}}^{(1)} e^{- i  k  t}~,
\end{eqnarray}\label{15a}
\end{subequations}
where the $\bm{k}$ is wave number, and the $\hat{k}$ describes propagation direction of the first order GWs. 

In the non-linear regime, we extend the calculations to the second-order perturbed geodesic equations, namely,
\begin{eqnarray}
	0 & = & \delta^2 P^{\mu} \partial_{\mu} P^{\nu} + 2 \delta P^{\mu}
	\partial_{\mu} \delta P^{\nu} + P^{\mu} \partial_{\mu} \delta^2 P^{\nu} + 2
	g^{\nu \rho} P^{\mu} \delta P^{\lambda} (\partial_{\mu} h^{(1)}_{\lambda \rho}
	- \partial_{\rho} h^{(1)}_{\mu \lambda} + \partial_{\lambda} h^{(1)}_{\mu \rho})
	\nonumber\\
	& &+ P^{\mu} P^{\lambda} \left( g^{\nu \rho} \left( \partial_{\lambda} \delta
	g^{(2)}_{\mu \rho} - \frac{1}{2} \partial_{\rho} \delta g_{\mu \lambda}^{(2)}
	\right) + g^{\nu \kappa} g^{\rho \omega} h^{(1)}_{\omega \kappa}
	(\partial_{\rho} h_{\mu \lambda}^{(1)} - 2 \partial_{\lambda} h^{(1)}_{\mu
			\rho}) \right)~, \label{15}
\end{eqnarray}
where where $\delta^2 P^\mu$ is the second order perturbed 4-momentum, and the second order metric perturbations in Synchronous gauge is
\begin{subequations}
\begin{eqnarray}
	\delta g^{(2)}_{0 0} & = & \delta g^{(2)}_{i 0} = 0~,\\
	\delta g^{(2)}_{i j} & = & - 2 \delta_{i  j} \psi^{(2)} + 2
	\partial_i \partial_j E^{(2)} + \partial_i C^{(2)}_j + \partial_j C_i^{(2)}
	+ h_{i  j}^{(2)}~.
\end{eqnarray}	\label{17}
\end{subequations}
To obtain the timing residuals in the second order, we calculate temporal components of Eq.~(\ref{15}) by making use of Eqs.~(\ref{11}) and (\ref{17}), namely,
\begin{eqnarray}
	(\partial_0 - \hat{n} \cdot \partial) \left( \frac{\delta^2 P^0}{P^0}
	\right) & = & - \frac{1}{2} \hat{n}^a \hat{n}^b \partial_0 \delta g_{a
			b}^{(2)} - 2 \left( \frac{\delta P^{\mu}}{P^0} \right)
	\partial_{\mu} \left( \frac{\delta P^0}{P^0} \right) + 2 \hat{n}^a \left(
	\frac{\delta P^b}{P^0} \right) \partial_0 h_{a  b}^{(1)}~.\label{18}
\end{eqnarray}
In contrast to the linear equations presented in Eq.~(\ref{14a}), the second and third terms on the right-hand side of the above equation represent non-linear contributions.
It was partly considered in previous study \cite{Tasinato:2022xyq}, in which the non-linear corrections were obtained by expanding the proper time to the second order. 
By making use of Eqs.~(\ref{15a}), the solutions of Eq.~(\ref{18}) can be obtained,
\begin{eqnarray}
	\frac{\delta^2 P^0_{\bm{k}}}{P^0} & = & \int \frac{{\rm d}^3 q}{(2
		\pi)^3} \left\{ \mathcal{F}^{c  d, a  b} \left( \bm{k},
	\bm{q} \right) \bar{h}_{c  d, \bm{k} - \bm{q}}^{(1)}
	\bar{h}^{(1)}_{a  b, \bm{q}} e^{- i \left( \left| \bm{k} -
			\bm{q} \right| + q \right) t} \right\}~, \label{19}
\end{eqnarray}
where
\begin{eqnarray}
		\mathcal{F}^{c  d, a  b} \left( \bm{k}, \bm{q},
	\hat{n} \right) & \equiv & - \frac{1}{\left| \bm{k} - \bm{q} \right| + q
		+ \hat{n} \cdot k} \left( \mathcal{F}_L^{c  d, a  b} \left(
	\bm{k}, \bm{q} \right) + \frac{\left| \bm{k} - \bm{q}
		\right| + q}{2} \sum_{\ast = \psi, E, C, h} \mathcal{F}^{c  d, a
		b}_{\ast} \left( \bm{k}, \bm{q} \right) \right)~, \nonumber\\
\end{eqnarray}
and the $\mathcal{F}_{**}^{c  d, a  b} \left(\bm{k}, \bm{q} \right)$, $(**=L,\psi, E, C, h)$ denote
\begin{subequations}
\begin{eqnarray}
	\mathcal{F}_L^{c  d, a  b} \left( \bm{k}, \bm{q}
	\right) & \equiv & \left( \left| \bm{k} - \bm{q} \right| + q \right)
	\mathcal{K}^{0, c  d}_{\bm{k - q}} \mathcal{K}^{0, a
		b}_{\bm{q}} - q_j \mathcal{K}^{j, c  d}_{\bm{k} -
	\bm{q}} \mathcal{K}^{0, a  b}_{\bm{q}}~, \nonumber\\
	 && - (k_j - q_j)
	\mathcal{K}^{0, c  d}_{\bm{k} - \bm{q}} \mathcal{K}^{j, a
		b}_{\bm{q}} - q \hat{n}^a \mathcal{K}^{b, c
		d}_{\bm{k} - \bm{q}} - \left| \bm{k} - \bm{q} \right|
	\hat{n}^c \mathcal{K}^{d, a  b}_{\bm{q}}~,\label{21a}\\
	\mathcal{F}^{c  d, a  b}_{\psi} \left( \bm{k},
	\bm{q} \right) & \equiv & - 2 \hat{F}_{\psi}^{i  j} \left(
	\bm{k}, \bm{q} \right) f_{i  j}^{c  d  a
			b} \left( \bm{k}, \bm{q} \right)~,\\
	\mathcal{F}^{c  d, a  b}_E \left( \bm{k}, \bm{q}
	\right) & \equiv & - 2 (\hat{n} \cdot k)^2 \hat{F}_E^{i  j} \left(
	\bm{k}, \bm{q} \right) f_{i  j}^{c  d  a
			b} \left( \bm{k}, \bm{q} \right)~,\\
	\mathcal{F}^{c  d, a  b}_C \left( \bm{k}, \bm{q}
	\right) & \equiv & i \hat{n}^j (\hat{n} \cdot k) \hat{F}^{m  l}_{C, j}
	\left( \bm{k}, \bm{q} \right) f_{m  l}^{c  d
			a  b} \left( \bm{k}, \bm{q} \right)~,\\
	\mathcal{F}_h^{c  d, a  b} \left( \bm{k}, \bm{q}
	\right) & \equiv & \hat{n}^i \hat{n}^j \hat{F}_{h, i  j}^{m  l}
	\left( \bm{k}, \bm{q} \right) f_{m  l}^{c  d
			a  b} \left( \bm{k}, \bm{q} \right)~.\label{21e}
\end{eqnarray}	
\end{subequations}
Remind that The ${\mathcal{K}}^{\mu,ab}_{\bm k}$ has been defined in Eq.~(\ref{15a}) and $\hat{F}_{\psi}^{i  j}\left( \bm{k}, \bm{q}\right)$, $\hat{F}_{E}^{i  j}\left( \bm{k}, \bm{q}\right)$, $\hat{F}_{C}^{i  j}\left( \bm{k}, \bm{q}\right)$ and $\hat{F}_{h}^{i  j}\left( \bm{k}, \bm{q}\right)$ have been given in Eq.~(\ref{7}).
The $\mathcal{F}_L^{c  d, a  b}\left( \bm{k}, \bm{q}\right)$ is derived from the second and the third terms of Eq.~(\ref{18}), and the $\mathcal{F}_\psi^{c  d, a  b}\left( \bm{k}, \bm{q}\right)$, $\mathcal{F}_E^{c  d, a  b}\left( \bm{k}, \bm{q}\right)$, $\mathcal{F}_C^{c  d, a  b}\left( \bm{k}, \bm{q}\right)$, and $\mathcal{F}_h^{c  d, a  b}\left( \bm{k}, \bm{q}\right)$ are obtained by the solving the motion of equations for the second order scalar, vector, and tensor perturbations, respectively. 

The difference of the time of arrivals can be quantified by the redshift caused by GWs fluctuations, namely, 
\begin{eqnarray}
	z & = & \frac{u_{\mu} \tilde{P}^{\mu} |_{\rm{obs}}}{ u_{\mu}
	\tilde{P}^{\mu} |_{\rm{src}}}~,
\end{eqnarray}
where $\tilde{P}^{\mu} (\equiv P^{\mu} + \delta P^{\mu} + \frac{1}{2} \delta^2	P^{\mu} +\mathcal{O} (3))$ is the total 4-momentum of a radio beam from a pulsar. For comoving observers in Synchronous gauge $u^{\mu} = (1, 0, 0, 0)$, we obtain the redshift and its fluctuations as follows,
\begin{subequations}
\begin{eqnarray}
	z^{(0)} & = & 0~,\\
	z^{(1)} & = & \frac{\delta P^0_{\rm{obs}}}{P^0_{\rm{obs}}} -
	\frac{\delta P^0_{\rm{src}}}{\delta P^0_{\rm{src}}}~,\\
	z^{(2)} & = & \frac{\delta^2 P^0_{\rm{obs}}}{P^0_{\rm{obs}}} -
	\frac{\delta^2 P^0_{\rm{src}}}{P^0_{\rm{src}}} - 2 \left( \frac{\delta
		P^0_{\rm{src}}}{P^0_{\rm{src}}} \right) \left( \frac{\delta
		P^0_{\rm{obs}}}{P^0_{\rm{obs}}} - \frac{\delta P^0_{\rm{src}}}{\delta
		P^0_{\rm{src}}} \right)~, 
\end{eqnarray}\label{23}
\end{subequations}
where the perturbed 4-momentums can be given by
$
	{\delta^{(n)} P^0}/{P^0} = \int \frac{{\rm d}^3 k}{(2 \pi)^3} \left\{
	({\delta^{(n)} P^0_{\bm{k}}}/{P^0}) e^{i  k \cdot x} \right\}
$,
and the expressions of $\delta^{(n)} P^0_{\bm{k}}/{P^0}$ have been given in Eqs.~(\ref{15a}) and (\ref{19}).
From Eq.~(\ref{11}), we have known the events of a pulsar emitting a radio beam $t_{\rm{src}} = t - L$ and $x^j_{\rm{src}} = L \hat{n}^j$, and the events of the beam reaching the earth $t_{\rm{obs}} = t$ and $x^j_{\rm{obs}} = 0$. Therefore, the redshifts and its fluctuations in Eqs.~(\ref{23}) can be rewritten in the form of
\begin{subequations}
\begin{eqnarray}
	z^{(1)} & = & \int \frac{{\rm d}^3 k}{(2 \pi)^3} \left\{ \mathcal{K}^{0, a
		b} \left( \bm{k}, \hat{n} \right) \bar{h}_{a  b,
	\bm{k}}^{(1)} e^{- i  k  t} (1 - e^{i  k
			L (1 + \hat{n} \cdot \hat{k})}) \right\}~, \label{25a}\\
	z^{(2)} & = & \int \frac{{\rm d}^3 k {\rm d}^3 q}{(2 \pi)^6} \left\{
	\mathcal{F}^{c  d, a  b} \left( \bm{k}, \bm{q},
	\hat{n} \right) \bar{h}_{c  d, \bm{k} - \bm{q}}^{(1)}
	\bar{h}_{a  b, \bm{q}}^{(1)} e^{- i  \left( \left|
			\bm{k} - \bm{q} \right| + q \right) t} \left( 1 - e^{i
		L \left( \left| \bm{k} - \bm{q} \right| + q +
		\hat{n} \cdot k \right)} \right)
	\right. \nonumber\\ && \left.
	+\mathcal{K}^{0, c  d} \left(
	\bm{k}, \hat{n} \right) \mathcal{K}^{0, a  b} \left( \bm{q},
	\hat{n} \right) \bar{h}_{c  d, \bm{k}}^{(1)} \bar{h}^{(1)}_{a
	b, \bm{q}} e^{- i  (k + q)  t} \right. \nonumber\\ && \left. \times \frac{1}{2}\left(e^{i  q
			L (1 + \hat{n} \cdot \hat{q})} (1 - e^{i  k  L (1 +
			\hat{n} \cdot \hat{k})})
			+e^{i  k
			L (1 + \hat{n} \cdot \hat{k})} (1 - e^{i  q  L (1 +
			\hat{n} \cdot \hat{q})})\right) \right\}~. \label{25b}
\end{eqnarray}\label{25}
\end{subequations}
In practice, the observables are the timing residuals obtained from pulsar timing measurements. It can be derived by integrating the redshifts over the observation duration $t$, namely, $R^{(n)}(t)=\int^t_0 z^{(n)}(\bar{t}) {\rm d} {\bar{t}}$, and
\begin{eqnarray}
	R^{(1)} & = & \int \frac{{\rm d}^3 k}{(2 \pi)^3} \left\{ \frac{1}{ik} \mathcal{K}^{0, a b} \left( \bm{k}, \hat{n} \right) \bar{h}_{a  b,
	\bm{k}}^{(1)} {(1-e^{- i  k  t})} (1 - e^{i  k
			L (1 + \hat{n} \cdot \hat{k})}) \right\}~,\\
	R^{(2)} & = & \int \frac{{\rm d}^3 k {\rm d}^3 q}{(2 \pi)^6} \left\{ \frac{1}{i\left( \left|
		\bm{k} - \bm{q} \right| + q \right)}
	\mathcal{F}^{c  d, a  b} \left( \bm{k}, \bm{q},
	\hat{n} \right) \bar{h}_{c  d, \bm{k} - \bm{q}}^{(1)}
	\bar{h}_{a  b, \bm{q}}^{(1)} (1-e^{- i  \left( \left|
	\bm{k} - \bm{q} \right| + q \right) t}) \left( 1 - e^{i
		L \left( \left| \bm{k} - \bm{q} \right| + q +
		\hat{n} \cdot k \right)} \right)
	\right. \nonumber\\ && \left.
	+ \frac{1}{i(k+q)}\mathcal{K}^{0, c  d} \left(
	\bm{k}, \hat{n} \right) \mathcal{K}^{0, a  b} \left( \bm{q},
	\hat{n} \right) \bar{h}_{c  d, \bm{k}}^{(1)} \bar{h}^{(1)}_{a
	b, \bm{q}} (1-e^{- i  (k + q)  t}) 
	\right. \nonumber\\ && \left. \times \frac{1}{2}
	\left(e^{i  q 
			L (1 + \hat{n} \cdot \hat{q})} (1 - e^{i  k  L (1 +
			\hat{n} \cdot \hat{k})})
			+e^{i  k
			L (1 + \hat{n} \cdot \hat{k})} (1 - e^{i  q  L (1 +
			\hat{n} \cdot \hat{q})})\right) \right\}~.
\end{eqnarray}
Due to $kt\ll 1$ for the nHz band PTAs, the timing residuals can be expressed in an expanded form as $kt \rightarrow 0$. By utilizing this approximation, the leading-order timing residuals reduce to $R^{(n)}=t z^{(n)}|_{t=0}$. It indicates that the corrections of the outputs (timing residuals) of PTAs can be obtained via the correlations of redshifts, namely, $\langle R(\bm x) R(\bm x') \rangle = t^2\langle z(\bm x) z(\bm x') \rangle $.

\section{Spatial correlations and overlap reduction functions}\label{IV}

\subsection{Non-linear correction of the correlations and non-Gaussianity}

Due to the stochastic nature of the SGWBs, the timing residuals should be studied statistically.
The spatial correlations of the timing residuals, as mentioned above, are proportional to spatial correlations of the redshifts $\langle z(\bm x) z(\bm x') \rangle$, where the $\bm x$ and $\bm x'$ can represent the locations of pulsar pairs. For illustration, we let $z_\alpha\equiv z(\bm x)$ and $z_\beta\equiv z(\bm x')$ in the rest of the paper. 

In order to compute the correlations of the redshifts, we expand them in the form of
\begin{eqnarray}
	\langle z_{\alpha} z_{\beta} \rangle & = & \langle z_{\alpha}^{(1)}
	z_{\beta}^{(1)} \rangle + \frac{1}{2} (\langle z_{\alpha}^{(1)}
	z_{\beta}^{(2)} \rangle + \langle z_{\alpha}^{(2)} z_{\beta}^{(1)} \rangle)
	+\mathcal{O} (4) ~.
\end{eqnarray}
For PTAs, the angular correlations derived from $\langle z_{\alpha}^{(1)} z_{\beta}^{(1)} \rangle$ is known as Hellings-Downs curve \cite{Hellings:1983fr}.
In this section, we will extend the spatial correlations to the non-linear regime with $\langle z_{\alpha}^{(1)} z_{\beta}^{(2)} \rangle$ and $\langle z_{\alpha}^{(2)} z_{\beta}^{(1)} \rangle$.
 
The purpose of PTAs is to extract the physical information of $h^{(1)}_{ij}$ from timing residuals of pulsar timing. Since $z^{(1)}\propto h^{(1)}$ and $z^{(2)}\propto (h^{(1)})^2$ shown in Eq.~(\ref{25}), the $\langle z_{\alpha}^{(1)} z_{\beta}^{(1)} \rangle$ and $\langle z_{\alpha}^{(1)} z_{\beta}^{(2)} \rangle$ could encode two-point and three-point correlations of $h_{ij}^{(1)}$, respectively. In this study, we confine our analysis to the isotropic and unpolarized GWs. In this case, the two-point correlations of $h^\lambda_{\bm k}$ in Fourier space can be given by
\begin{eqnarray}
	\left\langle h_{\bm{k}_1}^{(1), \lambda_1} h_{\bm{k}_2}^{(1),
			\lambda_2} \right\rangle & = & (2 \pi)^3 \delta \left( \bm{k}_1 +
	\bm{k}_2 \right) \delta^{\lambda_1 \lambda_2} P (k_2)~,\label{29}
\end{eqnarray}
where the $P(k)$ is power spectrum, the $\lambda_*(=+,\times)$ is polarization index, and the Kronecker symbol $\delta^{\lambda_1 \lambda_2}$ indicates that $h^{(1), \lambda}_{\bm{k}}$ is unpolarized. 
In the non-linear regime, the three-point correlations in Fourier space can be given by
\begin{eqnarray}
	\left\langle h_{\bm{k}_1}^{(1), \lambda_1} h_{\bm{k}_2}^{(1),
			\lambda_2} h_{\bm{k}_3}^{(1) , \lambda_3} \right\rangle & = & (2
	\pi)^6 \delta^{(3)} \left(
	\bm{k}_1 + \bm{k}_2 + \bm{k}_3 \right) \mathcal{B}^{\lambda_1\lambda_2\lambda_3}(k_1,k_2,k_3)~, \label{30}
\end{eqnarray}
where the $\mathcal{B}^{\lambda_1\lambda_2\lambda_3}(k_1,k_2,k_3)$ is bi-spectrum. It does not vanish due to the non-Gaussianity of SGWBs. From the non-polarization of the GWs, i) different polarizations of $h^{(1),\lambda}$ have no correlations, namely, $\mathcal{B}^{\lambda_1\lambda_2\lambda_3}(k_1,k_2,k_3)$ does not vanish, only if $\lambda_1=\lambda_2=\lambda_3$, and ii) different components of the bi-spectrum are equally weighted, namely, the $\mathcal{B}^{+++}(k_1,k_2,k_3)=\mathcal{B}^{\times\times\times}(k_1,k_2,k_3)$. 
To quantify shapes of the non-Gaussianity, we follow the parameterization scenario used in Ref.~\cite{Tasinato:2022xyq}. Finally, with all the above assumptions, the bi-spectrum in Eq.~(\ref{30}) reduces to
\begin{eqnarray}
	\mathcal{B}^{\lambda_1\lambda_2\lambda_3}(k_1,k_2,k_3)=H^{\lambda_1 \lambda_2 \lambda_3} (k_3) P (k_3) \delta (k_1 - \chi k_3)	\delta (k_2 - \zeta k_3)~, \label{31}
\end{eqnarray}
where we have $H^{+++}(k_3)=H^{\times\times\times}(k_3)\equiv \kappa(k_3)$ due to the non-polarization of GWs, the $P(k)$ is the power spectrum defined in Eq.~(\ref{29}), the $\zeta$ and $\chi$ are the dimensionless quantities formulating the shape of bi-spectrum, and here the $\kappa$ formulating the relative magnitude of bi-spectrum with respect to the power spectrum. 
The quantities $\kappa$, $\zeta$, and $\chi$ are, in principle, determined by the specific generation mechanism of GWs. Here, we did not involve any physical models for the generation mechanisms, but utilized the non-Gaussianity based on a phenomenological parameterization. 

\begin{figure}
	\includegraphics[width=.4\linewidth]{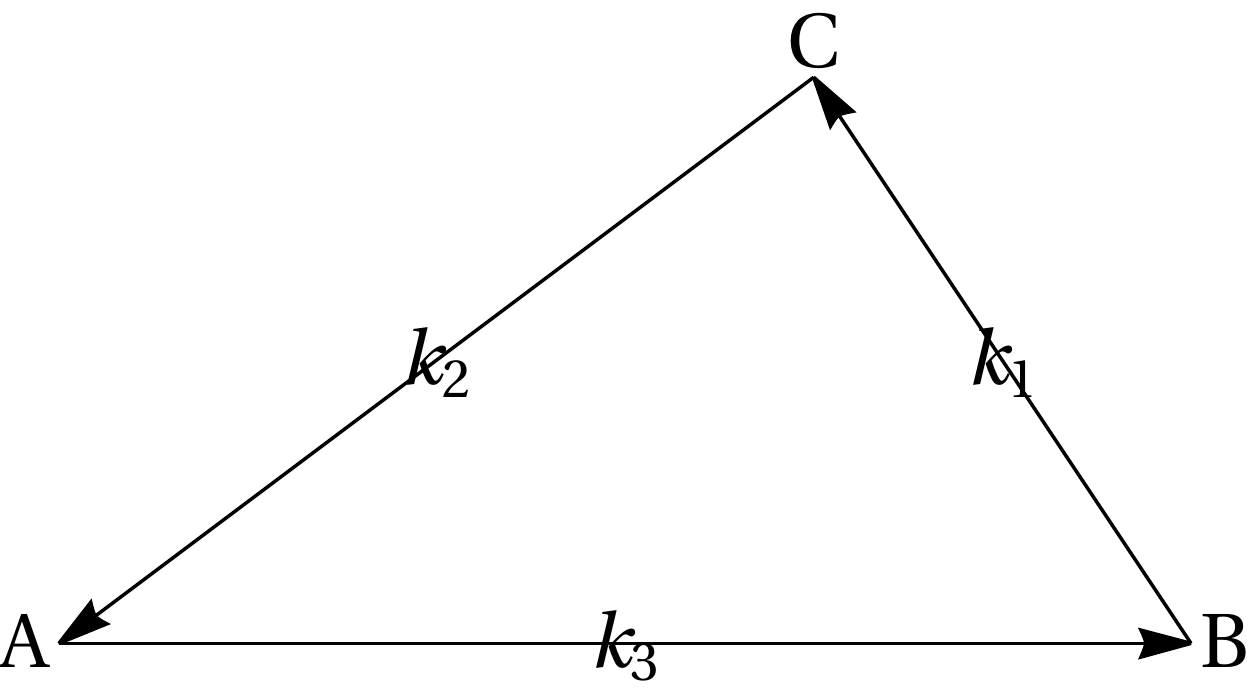}
	\caption{The scheme diagram of the parameterized bi-spectrum defined in Eq.~(\ref{31}). \label{F0}}
\end{figure}

\begin{figure}
	\includegraphics[width=1\linewidth]{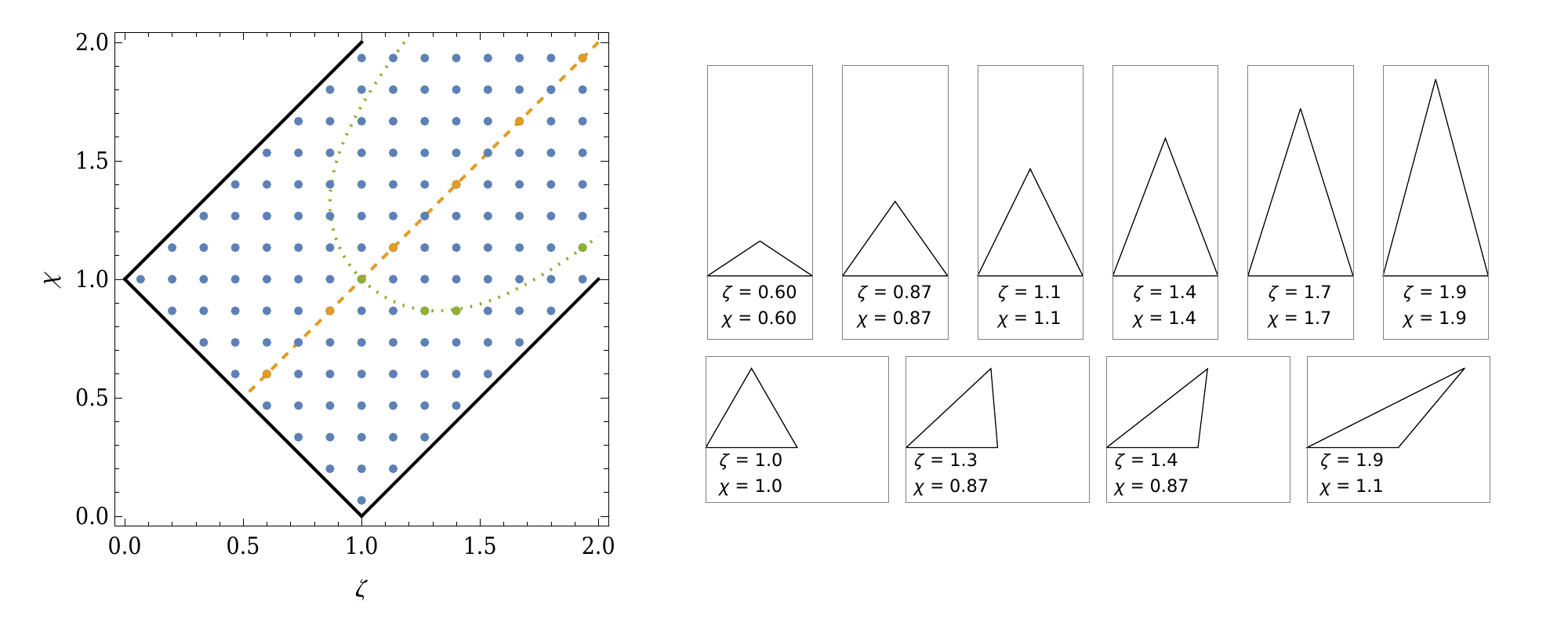}
	\caption{Left panel: parameter space of $(\zeta,\chi)$ for the parameterized non-Gaussianity of $h_{ij,k}^\lambda$. The points locate the available domain of the parameters. The dashed curve is formulated by $\zeta=\chi$, and the dotted curve is formulated by $\frac{1}{2}\zeta\chi\sqrt{1-\left(\frac{\zeta^2 + \chi^2 - 1}{2 \zeta \chi}\right)^2}=\frac{\sqrt{3}}{4}$. Right panel: The shape of non-Gaussianity for selected parameters $\zeta$ and $\chi$ on the dashed and dotted curves. \label{F2}}
\end{figure}

Because of $\bm{k}_1 + \bm{k}_2 + \bm{k}_3=0$ in Eq.~(\ref{30}), the parameters $\zeta$ and $\chi$ should satisfy the relations,
\begin{eqnarray}
	\chi + \zeta \geqslant 1~, \text{ and\ \ } |\chi - \zeta | \leqslant 1~. \label{32}
\end{eqnarray}
In Fig.~\ref{F0}, we show scheme diagram for the shape of the bi-spectrum, in which the $\zeta$ and $\chi$ are defined with $\zeta \equiv \overline{\rm BC}/\overline{\rm AB}$ and $\chi \equiv \overline{\rm CA}/\overline{\rm AB}$. In Fig.~\ref{F2}, we show the parameter space $(\zeta,\chi)$, where the points represent available parameters in Eqs.~(\ref{32}). Here, each individual point represents a distinct shape of non-Gaussianity. As shown in the right panel of Fig.~\ref{F2}, for example, the orange points on the dashed line represent isosceles triangles, and the green points on the dotted curve represent triangles with the same heights. 

Due to the central limit theorem, the SGWBs from the astrophysical origin are expected to be Gaussian \cite{Allen:1996vm}. In cosmology, the CMB observation suggested a Gaussian primordial curvature perturbation \cite{Planck:2019kim}. And it was also shown that influence from the non-Gaussianity of curvature perturbations could be erased at the late time \cite{DeLuca:2019qsy}. Thus, from the perspective of current experimental observation, the studies on the non-Gaussianity of SGWBs seem not to be very well-motivated. Merely, it would be of theoretical interest due to the non-linear nature of Einstein's gravity, which inevitably results in non-Gaussianity. In appendix~\ref{B}, we will provide an example of the non-Gaussianity of GWs arising from its non-linear generation.

\subsection{Overlap reduction functions of PTAs in second order}

By making use of Eqs.~(\ref{25a}) and (\ref{29}), the linear-order correlations of the redshifts for a pulsar pair can be given by
\begin{eqnarray}
	\langle z_{\alpha}^{(1)} z_{\beta}^{(1)} \rangle & = & \int \frac{k^2 {\rm d} k
	}{2 \pi^2} P (k) \int \frac{{\rm d} \Omega}{4 \pi} \left\{
	\mathcal{K}^{0, a  b} \left( \bm{k}, \hat{n}_{\alpha} \right)
	\mathcal{K}^{0, c  d} \left( \bm{k}, \hat{n}_{\beta} \right) 
	e^{\lambda}_{a  b} (\hat{k}) e^{\lambda}_{c  d} (\hat{k}) 
	\right. \nonumber \\	&& \left. \times
	(1	- e^{i  k  L_\alpha (1 + \hat{n}_{\alpha} \cdot \hat{k})}) (1 -
	e^{- i  k  L_\beta (1 + \hat{n}_{\beta} \cdot \hat{k})})
	\right\}\label{33}~,
\end{eqnarray}
where the $e^{\lambda}_{c  d} (\hat{k})$ is polarization tensor for $h^{(1)}_{ij, \bm{k}}$, and the $P (k)$ is the power spectrum defined in Eq.~(\ref{29}). 
The ORFs describe angular correlations of outputs of the GW detectors, and can be obtained by performing surface integrals over the unit sphere $\hat{k}$. Rewriting Eq.~(\ref{33}) in the form of
\begin{eqnarray}
	\langle z_{\alpha}^{(1)} z_{\beta}^{(1)} \rangle 
	& \equiv & \int \frac{k^2 {\rm d} k}{2 \pi^2}  P (k)	\Gamma^{\rm{(2)}} (k, \theta_{\alpha \beta}) ~,
\end{eqnarray}
one can read the ORFs,
\begin{eqnarray}
	\Gamma^{\rm{(2)}} (k, \theta_{a  b}) &=& \int \frac{{\rm d} \Omega}{4 \pi} \left\{
	\mathcal{K}^{0, a  b} \left( \bm{k}, \hat{n}_{\alpha} \right)
	\mathcal{K}^{0, c  d} \left( \bm{k}, \hat{n}_{\beta} \right)
	e^{\lambda}_{a  b} (\hat{k}) e^{\lambda}_{c  d} (\hat{k}) 
		\right. \nonumber \\	&&\left. \times
		(1	- e^{i  k  L_\alpha (1 + \hat{n}_{\alpha} \cdot \hat{k})}) (1 -
	e^{- i  k  L_\beta (1 + \hat{n}_{\beta} \cdot \hat{k})})
	\right\}~, \label{35}
\end{eqnarray}
where $\theta_{\alpha \beta} \equiv \cos^{- 1}	(\hat{n}_{\alpha} \cdot \hat{n}_{\beta})$.
Because of the approximation $kL\gg1$ with the known frequency band and arm's length of PTAs, the above ORFs would reduce to Hellings-Downs curve \cite{Hellings:1983fr}, namely,
\begin{eqnarray}
	\Gamma^{\rm{HD}} (\theta_{a  b}) \equiv \Gamma^{\rm{(2)}} (k, \theta_{a  b})|_{kL_\alpha\gg1,kL_\beta\gg1} = \int \frac{{\rm d} \Omega}{4 \pi} \left\{
	\mathcal{K}^{0, a  b} \left( \bm{k}, \hat{n}_{\alpha} \right)
	\mathcal{K}^{0, c  d} \left( \bm{k}, \hat{n}_{\beta} \right)
	e^{\lambda}_{a  b} (\hat{k}) e^{\lambda}_{c  d} (\hat{k})	\right\}~.
\end{eqnarray} 
Here, the oscillation parts in Eq.~(\ref{35}) is suppressed by the factor $(kL)^{-1}$, and thus can be neglected for PTAs. Besides, the ORFs without the approximation were also studied \cite{Mingarelli:2014xfa,Boitier:2020rzg,Hu:2022ujx}. 

In the second order, we further compute non-linear corrections
for the correlations of redshift in Eq.~(\ref{25}), namely,
\begin{eqnarray}
	\langle z^{(1)}_{\alpha} z^{(2)}_{\beta} \rangle & = & \int \frac{{\rm d}^3 k
		{\rm d}^3 k' {\rm d}^3 q'}{(2 \pi)^{12}} \left\{ \left\langle h^{
			(1)}_{\bm{k}, m  l} {h^{(1)}_{\bm{k}' - \bm{q}', c
			d}}^{*} {h^{(1)}_{\bm{q}', a  b}}^{*}
	\right\rangle \mathcal{K}^{0, m  l} \left( \bm{k},
	\hat{n}_{\alpha} \right) \mathcal{F}^{c  d, a  b} \left(
	\bm{k}', \bm{q}', \hat{n}_{\beta} \right)
	\right. \nonumber\\ && \left. \times
	e^{- i  k
			t} (1 - e^{i  k  L_\alpha(1 + \hat{n}_{\alpha} \cdot \hat{k})})
	e^{i  t \left( \left| \bm{k}' - \bm{q}' \right| + q'
			\right)} \left( 1 - e^{- i  L_\beta\left( \left| \bm{k}' -
		\bm{q}' \right| + q' + \hat{n}_{\beta} \cdot k' \right)} \right)  \right. \nonumber\\ && \left. +
	\frac{1}{2} \left\langle h^{ (1)}_{\bm{k}, m  l} {h^{(1)}_{\bm{k}', c
			d}}^{*} {h^{(1)}_{\bm{q}', a  b}}^{*}
	\right\rangle \mathcal{K}^{0, m  l} \left( \bm{k},
	\hat{n}_{\alpha} \right) \mathcal{K}^{0, c  d} \left( \bm{k}',
	\hat{n}_{\beta} \right) \mathcal{K}^{0, a  b} \left( \bm{q}',
	\hat{n}_{\beta} \right) e^{- i  k  t} (1 - e^{i  k
			L_\alpha(1 + \hat{n}_{\alpha} \cdot \hat{k})}) e^{i (k' + q') t} 
			\right. \nonumber\\ && \left. \times 
			\left(e^{- i
			q' L_\beta(1 + \hat{n}_{\beta} \cdot \hat{q}')} (1 - e^{- i  k'
			L_\beta(1 + \hat{n}_{\beta} \cdot \hat{k}')})+e^{- i
			q' L_\beta(1 + \hat{n}_{\beta} \cdot \hat{q}')} (1 - e^{- i  k'
			L_\beta(1 + \hat{n}_{\beta} \cdot \hat{k}')})\right) \right\}~,\label{37}
\end{eqnarray}
where above three-point correlations of $h^{(1)}_{i  j, \bm{k}}$ can be written
in terms of polarization components $h^{(1), \lambda}_{\bm{k}}$, 
\begin{eqnarray}
	\left\langle h_{\bm{k}, c  d}^{(1)} h_{a  b,
			\bm{p}}^{(1)} h^{(1)}_{\bm{q}, m  l} \right\rangle & = &
	e^{\lambda_1}_{c  d} (\hat{k}) e^{\lambda_2}_{a  b}
	(\hat{p}) e^{\lambda_3}_{m  l} (\hat{q}) \left\langle h^{(1),
			\lambda_1}_{\bm{k}} h^{(1), \lambda_2}_{\bm{p}} h^{(1),
			\lambda_3}_{\bm{q}} \right\rangle~.
\end{eqnarray}
By making use of Eqs.~(\ref{30}) and (\ref{31}), the Eq.~(\ref{35}) is evaluated to be
\begin{eqnarray}
	\langle z^{(1)}_{\alpha} z^{(2)}_{\beta} \rangle + \langle z_{\alpha}^{(2)}
	z_{\beta}^{(1)} \rangle & = & \int \frac{{\rm d}^3 k}{(2 \pi)^3} \int
	\frac{{\rm d} \phi_q}{2 \pi} \left\{ e^{\lambda_1}_{m  l} (\hat{k})
	e^{\lambda_2}_{c  d} (\widehat{k - q}) e^{\lambda_3}_{a  b}
	(\hat{q}) H^{\lambda_1 \lambda_2 \lambda_3} (k) P (k)
	\right. \nonumber\\ && \left.
	\times \left( \left(
	\mathcal{K}^{0, m  l} \left( \bm{k}, \hat{n}_{\alpha} \right)
	\mathcal{F}^{c  d, a  b} \left( \bm{k}, \bm{q},
	\hat{n}_{\beta} \right) (1 - e^{i  k  L_\alpha(1 +
				\hat{n}_{\alpha} \cdot \hat{k})}) (1 - e^{- i  k  L_\beta(\chi +
				\zeta + \hat{n}_{\beta} \cdot \hat{k})})
	\right.\right. \right. \nonumber\\ && \left.\left.\left.
	+\frac{1}{2}\mathcal{K}^{0, m  l}
	\left( \bm{k}, \hat{n}_{\alpha} \right) \mathcal{K}^{0, c  d}
	\left( \bm{k} - \bm{q}, \hat{n}_{\beta} \right) \mathcal{K}^{0, a
		b} \left( \bm{q}, \hat{n}_{\beta} \right)
	\right.\right. \right. \nonumber\\ && \left.\left.\left.
	\times e^{- i  \zeta k
			L_\beta(1 + \hat{n}_{\beta} \cdot \hat{q})} (1 - e^{i  k
			L_\alpha(1 + \hat{n}_{\alpha} \cdot \hat{k})}) (1 - e^{- i  L_\beta
			(\chi    k + \hat{n}_{\beta} \cdot (k - q))}) 
	\right.\right. \right. \nonumber\\ && \left.\left.\left.
	+\frac{1}{2}\mathcal{K}^{0, m  l}
	\left( \bm{k}, \hat{n}_{\alpha} \right) \mathcal{K}^{0, c  d}
	\left( \bm{q}, \hat{n}_{\beta} \right) \mathcal{K}^{0, a
		b} \left(\bm{k} - \bm{q}, \hat{n}_{\beta} \right)
	\right.\right. \right. \nonumber\\ && \left.\left.\left.
	\times e^{- i  
			L_\beta(\chi k + \hat{n}_{\beta} \cdot {(k-q)})} (1 - e^{i  k
			L_\alpha(1 + \hat{n}_{\alpha} \cdot \hat{k})}) (1 - e^{- i \zeta k L_\beta
			(1 + \hat{n}_{\beta} \cdot \hat{q}))}) \right) e^{- i
			k  t (1 - \chi - \zeta)}
	\right.\right. \nonumber\\ && \left.\left.
	+ \left( \mathcal{F}^{c
		d, a  b} \left( \bm{k}, \bm{q}, \hat{n}_{\alpha} \right)
	\mathcal{K}^{0, m  l} \left( \bm{k}, \hat{n}_{\beta} \right) (1
	- e^{i  k  L_\alpha(\zeta + \chi + \hat{n}_{\alpha} \cdot
				\hat{k})}) (1 - e^{- i  k  L_\beta(1 + \hat{n}_{\beta} \cdot
				\hat{k})})
	\right.\right. \right. \nonumber\\ && \left.\left.\left.
	+\frac{1}{2}\mathcal{K}^{m  l} \left( \bm{k}, \hat{n}_{\beta}
	\right) \mathcal{K}^{0, c  d} \left( \bm{k} - \bm{q},
	\hat{n}_{\alpha} \right) \mathcal{K}^{ 0, a  b} \left(
	\bm{q}, \hat{n}_{\alpha} \right)
	\right.\right. \right. \nonumber\\ && \left.\left.\left.
	\times e^{i  \zeta k  L_\alpha(1 +
			\hat{n}_{\alpha} \cdot \hat{q})} (1 - e^{i  L_\alpha (\chi k +
			\hat{n}_{\alpha} \cdot (k - q))}) (1 - e^{- i  k  L_\beta(1 +
			\hat{n}_{\beta} \cdot \hat{k})})
	\right.\right. \right. \nonumber\\ && \left.\left.\left.
	+\frac{1}{2}\mathcal{K}^{m  l} \left( \bm{k}, \hat{n}_{\beta}
	\right) \mathcal{K}^{0, c  d} \left( \bm{q} ,
	\hat{n}_{\alpha} \right) \mathcal{K}^{ 0, a  b} \left(
	\bm{k} - \bm{q}, \hat{n}_{\alpha} \right)
	\right.\right. \right. \nonumber\\ && \left.\left.\left.
	\times e^{i     L_\alpha(\chi k +
			\hat{n}_{\alpha} \cdot (k-q))} (1 - e^{i \zeta k L_\alpha ( 1+
			\hat{n}_{\alpha} \cdot \hat{q})}) (1 - e^{- i  k  L_\beta(1 +
			\hat{n}_{\beta} \cdot \hat{k})}) \right) e^{i  k  t (1 -
			\chi - \zeta)} \right) \right\}~, \nonumber\\ \label{40}
\end{eqnarray}
where the momentums are given by
\begin{subequations}
\begin{eqnarray}
	\bm{q} & = & \left( \frac{1 + \zeta^2 - \chi^2}{2} \right) \bm{k} +
	\frac{\sqrt{((\zeta + \chi)^2 - 1) (1 - (\zeta - \chi)^2)}}{2} k \left( \cos
	\phi_q \bm{u} + \sin \phi_q \bm{\upsilon} \right)~,\label{41a}\\
	\hat{q} & = & \frac{1 + \zeta^2 - \chi^2}{2 \zeta} \hat{k} +
	\frac{\sqrt{((\zeta + \chi)^2 - 1) (1 - (\zeta - \chi)^2)}}{2 \zeta} \left(
	\cos \phi_q \bm{u} + \sin \phi_q \bm{\upsilon} \right)~,\\
	\widehat{k - q} & = & \frac{\chi^2 - \zeta^2 + 1}{2 \chi} \hat{k} -
	\frac{\sqrt{((\zeta + \chi)^2 - 1) (1 - (\zeta - \chi)^2)}}{2 \chi} \left(
	\cos \phi_q \bm{u} + \sin \phi_q \bm{\upsilon} \right)~.
\end{eqnarray}	
\end{subequations}
Here, the $\bm{u}$ and $\bm{\upsilon}$ represent polarization vectors	with respect to the $\hat{k}$. Using Eq.~(\ref{41a}), one can verify the relations of $q=\zeta k$ and $|{\bm k} - {\bm q}| = \chi k$ shown in Fig.~\ref{F0}. From Eq.~(\ref{40}), it is found that the three-point correlations are proportional to $e^{\pm i  k  t (1 - \chi - \zeta)}$. It indicates that the values of correlations would oscillate with time around zero. In practice, due to $k  t \ll 1$ in nHz band of PTAs, we here can let $e^{\pm i  k  t (1 - \chi - \zeta)} \simeq	1$.

Similarly, the correlations in Eq.~(\ref{40}) can be rewritten in the form of
\begin{eqnarray}
		\langle z^{(1)}_{\alpha} z^{(2)}_{\beta} \rangle + \langle z_{\alpha}^{(2)}
	z_{\beta}^{(1)} \rangle = \int \frac{k^2 {\rm d} k}{2 \pi^2} P (k)
	\Gamma^{(3)} (k, \theta_{a  b})~, 
\end{eqnarray}
where the ORFs in the non-linear order are given by
\begin{eqnarray}
	\Gamma^{(3)} (k, \theta_{\alpha \beta}) & = & \int \frac{{\rm d}
		\Omega}{4 \pi} \int
	\frac{{\rm d} \phi_q}{2 \pi} \left\{ e^{\lambda_1}_{m  l} (\hat{k})
	e^{\lambda_2}_{c  d} (\widehat{k - q}) e^{\lambda_3}_{a  b}
	(\hat{q}) H^{\lambda_1 \lambda_2 \lambda_3} (k) P (k)
	\right. \nonumber\\ && \left.
	\times \left( 
	\mathcal{K}^{0, m  l} \left( \bm{k}, \hat{n}_{\alpha} \right)
	\mathcal{F}^{c  d, a  b} \left( \bm{k}, \bm{q},
	\hat{n}_{\beta} \right) (1 - e^{i  k  L_\alpha(1 +
				\hat{n}_{\alpha} \cdot \hat{k})}) (1 - e^{- i  k  L_\beta(\chi +
				\zeta + \hat{n}_{\beta} \cdot \hat{k})})
	\right.\right. \nonumber\\ && \left.\left.
	+ \mathcal{F}^{c
		d, a  b} \left( \bm{k}, \bm{q}, \hat{n}_{\alpha} \right)
	\mathcal{K}^{0, m  l} \left( \bm{k}, \hat{n}_{\beta} \right) (1
	- e^{i  k  L_\alpha(\zeta + \chi + \hat{n}_{\alpha} \cdot
				\hat{k})}) (1 - e^{- i  k  L_\beta(1 + \hat{n}_{\beta} \cdot
				\hat{k})})
	\right. \right. \nonumber\\ && \left.\left.
	+\frac{1}{2}\mathcal{K}^{0, m  l}
	\left( \bm{k}, \hat{n}_{\alpha} \right) \mathcal{K}^{0, c  d}
	\left( \bm{k} - \bm{q}, \hat{n}_{\beta} \right) \mathcal{K}^{0, a
		b} \left( \bm{q}, \hat{n}_{\beta} \right)
	\right. \right. \nonumber\\ && \left.\left.
	\times e^{- i  \zeta k
			L_\beta(1 + \hat{n}_{\beta} \cdot \hat{q})} (1 - e^{i  k
			L_\alpha(1 + \hat{n}_{\alpha} \cdot \hat{k})}) (1 - e^{- i  L_\beta
			(\chi    k + \hat{n}_{\beta} \cdot (k - q))}) 
	\right. \right. \nonumber\\ && \left.\left.
	+\frac{1}{2}\mathcal{K}^{0, m  l}
	\left( \bm{k}, \hat{n}_{\alpha} \right) \mathcal{K}^{0, c  d}
	\left( \bm{q}, \hat{n}_{\beta} \right) \mathcal{K}^{0, a
		b} \left(\bm{k} - \bm{q}, \hat{n}_{\beta} \right)
	\right. \right. \nonumber\\ && \left.\left.
	\times e^{- i  
			L_\beta(\chi k + \hat{n}_{\beta} \cdot {(k-q)})} (1 - e^{i  k
			L_\alpha(1 + \hat{n}_{\alpha} \cdot \hat{k})}) (1 - e^{- i \zeta k L_\beta
			(1 + \hat{n}_{\beta} \cdot \hat{q}))}) 
	\right. \right. \nonumber\\ && \left.\left.
	+\frac{1}{2}\mathcal{K}^{m  l} \left( \bm{k}, \hat{n}_{\beta}
	\right) \mathcal{K}^{0, c  d} \left( \bm{k} - \bm{q},
	\hat{n}_{\alpha} \right) \mathcal{K}^{ 0, a  b} \left(
	\bm{q}, \hat{n}_{\alpha} \right)
	\right. \right. \nonumber\\ && \left.\left.
	\times e^{i  \zeta k  L_\alpha(1 +
			\hat{n}_{\alpha} \cdot \hat{q})} (1 - e^{i  L_\alpha (\chi k +
			\hat{n}_{\alpha} \cdot (k - q))}) (1 - e^{- i  k  L_\beta(1 +
			\hat{n}_{\beta} \cdot \hat{k})})
	\right. \right. \nonumber\\ && \left.\left.
	+\frac{1}{2}\mathcal{K}^{m  l} \left( \bm{k}, \hat{n}_{\beta}
	\right) \mathcal{K}^{0, c  d} \left( \bm{q} ,
	\hat{n}_{\alpha} \right) \mathcal{K}^{ 0, a  b} \left(
	\bm{k} - \bm{q}, \hat{n}_{\alpha} \right)
	\right. \right. \nonumber\\ && \left.\left.
	\times e^{i     L_\alpha(\chi k +
			\hat{n}_{\alpha} \cdot (k-q))} (1 - e^{i \zeta k L_\alpha ( 1+
			\hat{n}_{\alpha} \cdot \hat{q})}) (1 - e^{- i  k  L_\beta(1 +
			\hat{n}_{\beta} \cdot \hat{k})}) \right) \right\}~. \label{42}
\end{eqnarray}
The expression of $H^{\lambda_1 \lambda_2 \lambda_3} (k)$ has been given in the Eq.~(\ref{31}).
Since the oscillation parts in the integration are suppressed by the factor $(kL)^{-1}$ for PTAs, we also adopt $kL_\alpha\gg1$ and $kL_\beta\gg1$ for evaluating Eq.~(\ref{42}). Namely, the ORFs can be simplified in the form of
\begin{eqnarray}
	\Gamma^{\rm{nl}} (k, \theta_{\alpha \beta}) &\equiv& \Gamma^{\rm{(3)}} (k, \theta_{a  b})|_{kL_\alpha\gg1,kL_\beta\gg1}\nonumber\\
	 & = & \kappa\int \frac{{\rm d}
		\Omega}{4 \pi} \int \frac{{\rm d} \phi_q}{2 \pi} \left\{ \left(e^+_{m  l} (\hat{k}) e^+_{c  d}
	(\widehat{k - q}) e^+_{a  b} (\hat{q}) + e^{\times}_{m  l}
	(\hat{k}) e^{\times}_{c  d} (\widehat{k - q}) e^{\times}_{a
			b} (\hat{q})\right)
		\right. \nonumber\\ && \left. \times
	\left( \mathcal{K}^{0, m  l} \left( \bm{k}, \hat{n}_{\alpha}
		\right) \mathcal{F}^{c  d, a  b} \left( \bm{k},
		\bm{q}, \hat{n}_{\beta} \right)
	+\mathcal{F}^{c  d, a b} \left( \bm{k}, \bm{q}, \hat{n}_{\alpha} \right) \mathcal{K}^{0, m l} \left( \bm{k}, \hat{n}_{\beta} \right) \right) \right\}~. \label{43}
\end{eqnarray}

\begin{figure}
	\includegraphics[width=1\linewidth]{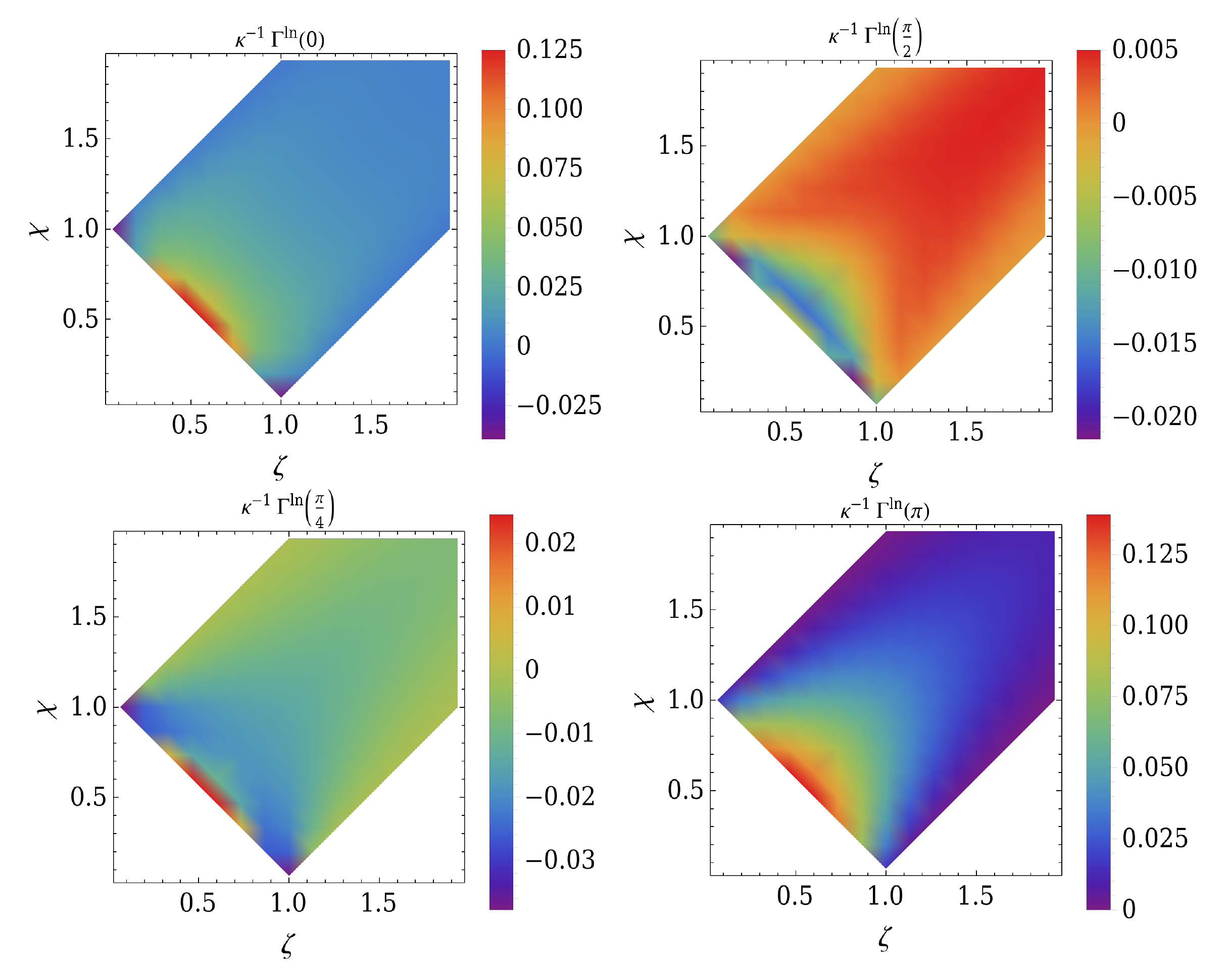}
	\caption{Non-linear ORFs over the $\kappa$ as function of parameter $(\zeta,\chi)$ for $\theta_{\alpha\beta}=0,\pi/4,\pi/2,\pi$. \label{F3}}
\end{figure}

\begin{figure}
	\includegraphics[width=0.8\linewidth]{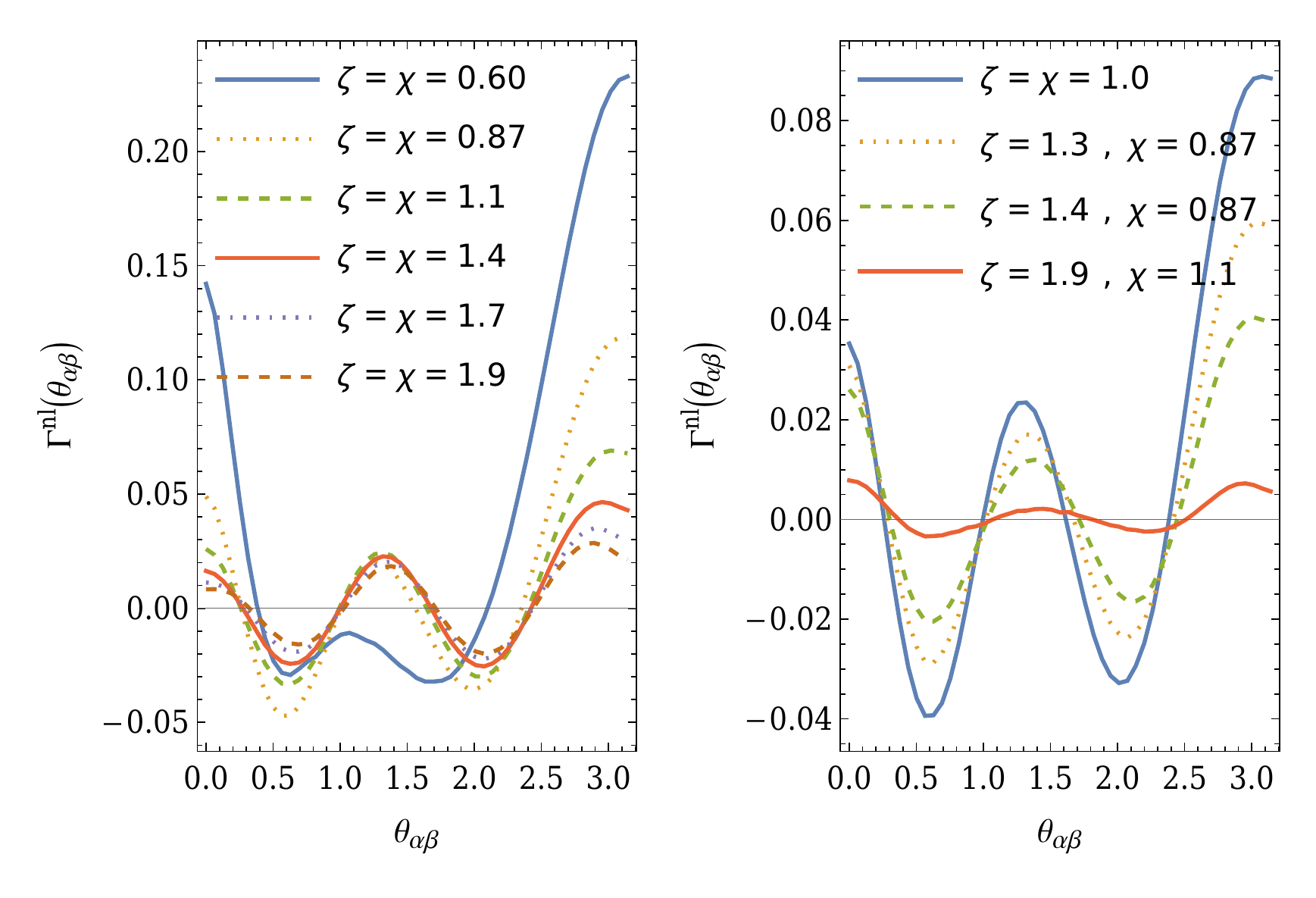}
	\caption{Non-linear ORFs for selected parameters $\zeta$ and $\chi$ shown in right panel of Fig.~\ref{F2}\label{F4}, and $\kappa=1$. Left panel: the non-Gaussianity in the shape of isosceles triangles. Right panel: the non-Gaussianity in the shape of the triangles with the same height.}
\end{figure}

\begin{figure}
	\includegraphics[width=0.8\linewidth]{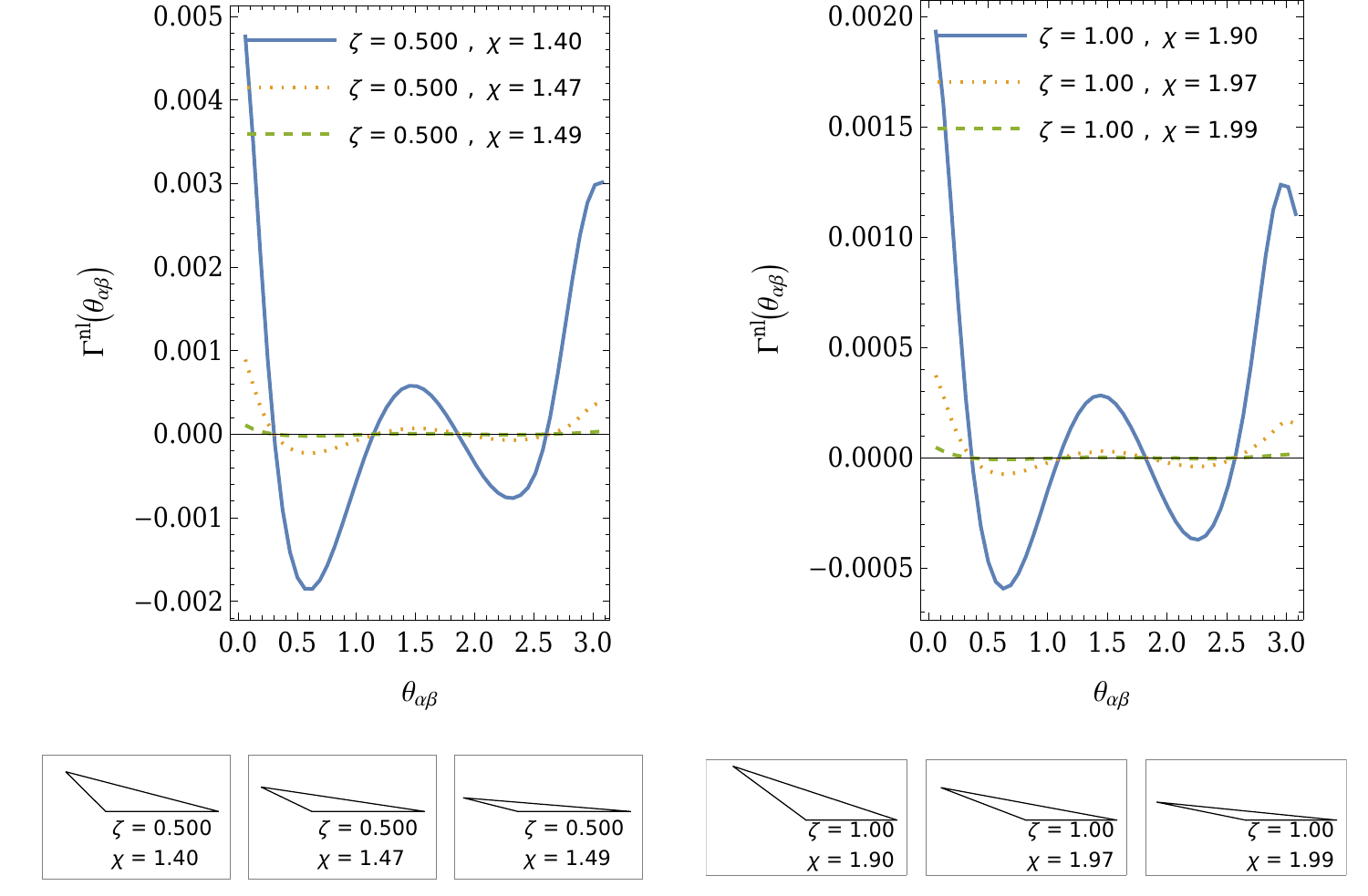}
	\caption{Non-linear ORFs for selected parameters $\zeta-\chi\rightarrow1$, and $\kappa=1$. \label{F5}}
\end{figure}

\begin{figure}
	\includegraphics[width=0.5\linewidth]{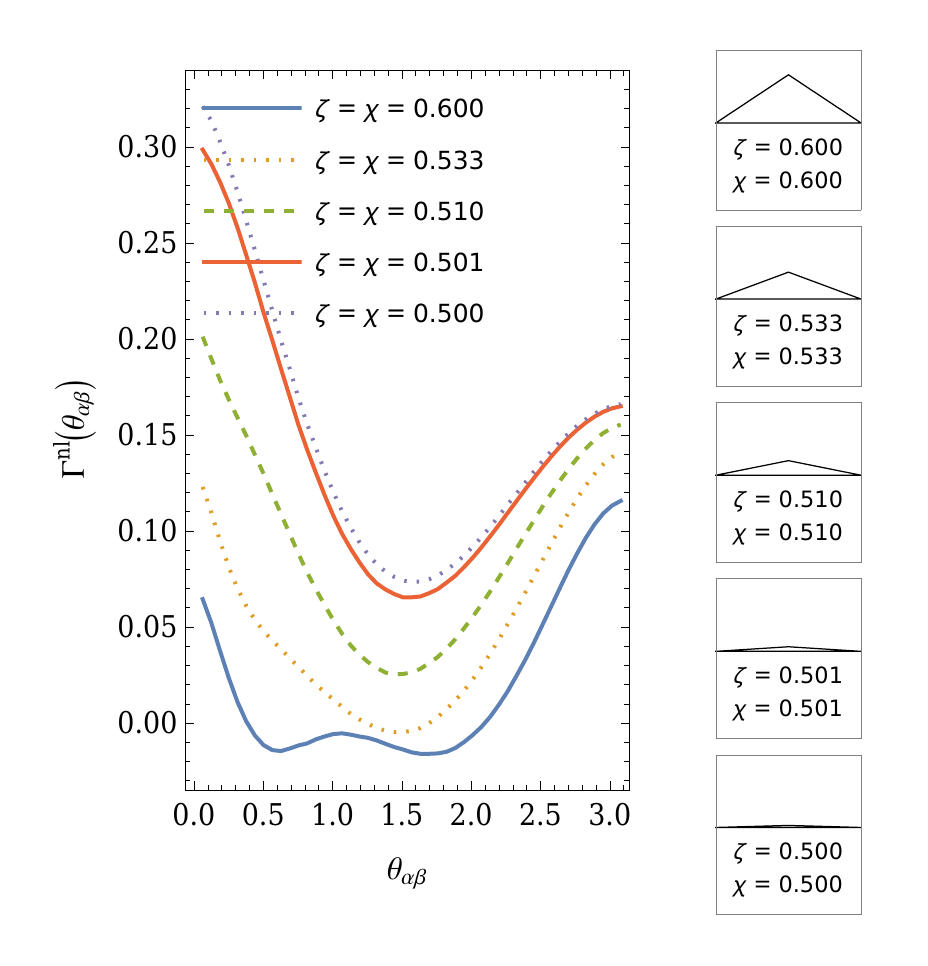}
	\caption{Non-linear ORFs for selected parameters $\zeta+\chi\rightarrow1$, and $\kappa=1$. \label{F6}}
\end{figure}

In the following, we will present the results of $\Gamma^{\rm{nl}} (k,\theta_{a  b})$ with selected parameters $\zeta$ and $\chi$.
In Fig.~\ref{F3}, it shows the ORFs over the $\kappa$ as function of parameter $(\zeta,\chi)$ for given angle $\theta_{\alpha \beta}$. It is found that the values of $\Gamma^{\rm ln}/\kappa$ tend to approach zero for larger values of $\zeta$ and $\chi$, and exhibit their largest magnitude when $\zeta+\chi=1$.
In Fig.~\ref{F4}, we show the ORFs for select parameters given in the right panel of Fig.~\ref{F2}. 
It consistently shows that as the values of $\zeta$ and $\chi$ increase, the ratio of $\Gamma^{\rm ln}$ to $\kappa$ tends to decrease.
In contrast to the ORFs in the linear order, the curves of the non-linear correction of ORFs exhibit three distinct extreme points. Besides, in order to clarify the extreme cases, such as $\zeta-\chi=1$ or $\zeta+\chi=1$, we show the ORFs as function of $\theta_{\alpha\beta}$ for $\zeta-\chi\rightarrow1$ and $\zeta+\chi\rightarrow1$ in Figs.~\ref{F5} and \ref{F6}, respectively. From Fig.~\ref{F5}, the values of ORFs tend to vanish as $\zeta-\chi\rightarrow1$, and these ORFs have the same zero points with respect to $\theta_{\alpha\beta}$. From Fig.~\ref{F6}, the values of ORFs tend to be larger, and the numbers of extreme points are described in the case of $\zeta+\chi\rightarrow1$. It is different from the results shown in the left panel of Fig.~\ref{F4} for a larger $\zeta+\chi$. 

\begin{figure}
	\includegraphics[width=0.5\linewidth]{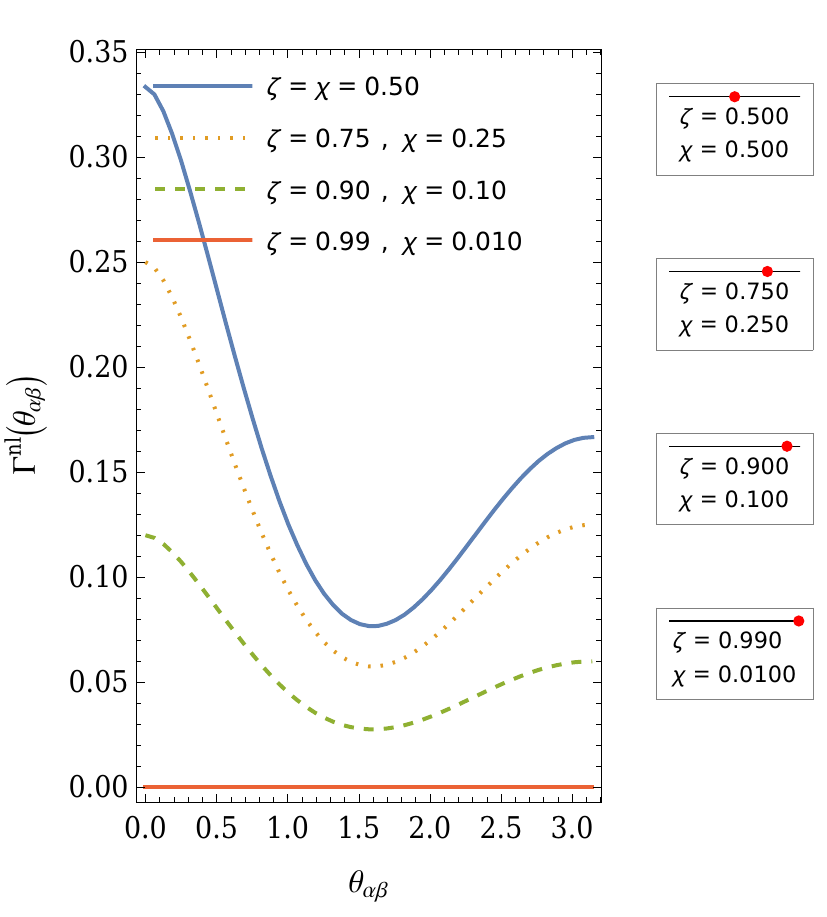}
	\caption{Non-linear ORFs for selected parameters $\zeta+\chi=1$, $\zeta\rightarrow1$, and $\kappa=1$. \label{F7}} 
\end{figure}

For the shape of non-Gaussianity being a straight line with $\zeta + \chi = 1$, we can further simplify the expression of ORFs as
\begin{eqnarray}
	\Gamma^{\rm{nl}} (k, \theta_{\alpha \beta}) & = & \kappa \int \frac{{\rm d}
		\Omega}{4 \pi} \left\{ \left(e^+_{m  l} (\hat{k}) e^+_{c  d}
	(\widehat{k}) e^+_{a  b} (\hat{k}) + e^{\times}_{m  l}
	(\hat{k}) e^{\times}_{c  d} (\widehat{k}) e^{\times}_{a
			b} (\hat{k})\right) \right. \nonumber\\ 
	&& \left. \times \left(
	\mathcal{K}^{0, m  l} \left( \bm{k}, \hat{n}_{\alpha} \right)
	\mathcal{F}^{c  d, a  b} \left( \bm{k}, \zeta
	\bm{k}, \hat{n}_{\beta} \right)
	+\mathcal{F}^{c  d, a  b} \left( \bm{k}, \zeta
	\bm{k}, \hat{n}_{\alpha} \right) \mathcal{K}^{0, m  l} \left(
	\bm{k}, \hat{n}_{\beta} \right) \right) \right\}~.
\end{eqnarray}
Here, the integration over the angle $\phi_q$ simply gives $2 \pi$. In Fig.~\ref{F7}, we show the ORFs in the case of $\zeta+\chi=1$ for different $\zeta$. It is found that the values of ORFs get smaller as $\zeta\rightarrow0$. 

\begin{figure} 
	\includegraphics[width=0.7\linewidth]{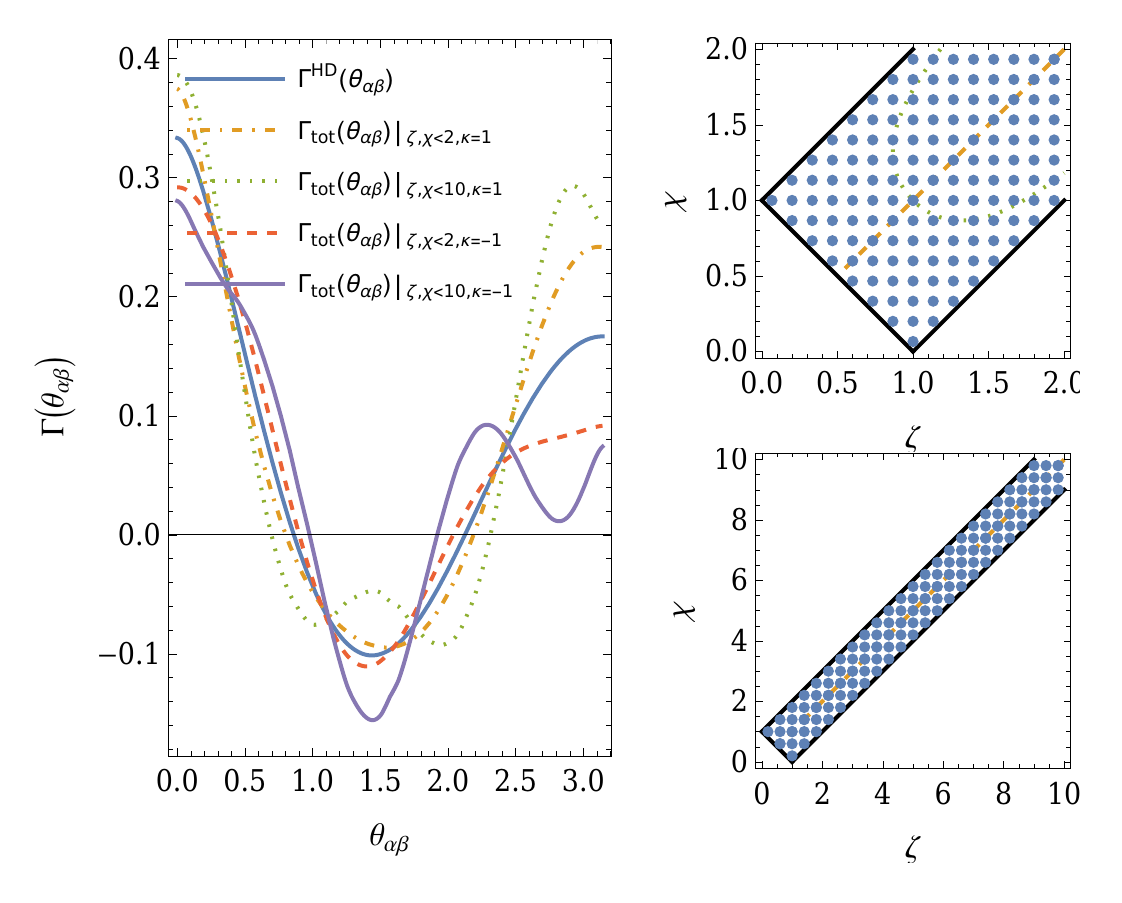}
	\caption{Left panel: total ORFs with non-linear corrections. The non-linear ORFs are the sum of the selected parameters $\kappa=\pm1$ and $(\zeta,\chi)$ shown with the points in the right panel. Right panel: The sets of selected parameters in the plots of $(\zeta,\chi)$. The $\zeta,\chi\in (0,2)$, and $\zeta,\chi\in (0,10)$ for the top-right panel, and bottom-right panel, respectively. \label{F8}} 
\end{figure}

The parameter $\kappa$ in Eq.~(\ref{43}) is dependent on the wave number $k$, as well as the shape of non-Gaussianity as quantified by $\zeta$ and $\chi$.
There seems no reason that the non-linear corrections come from the non-Gaussianity with one of the available parameters $(\zeta,\chi)$. Therefore, in theory, the total ORFs to the non-linear regime should be the sum of all the shapes of non-Gaussianity weighted by parameter $\kappa$, namely,
\begin{eqnarray}
	\Gamma (k, \theta_{\alpha \beta}) & = & \Gamma^{\rm{HD}} (
	\theta_{\alpha \beta}) + \frac{1}{2} \sum_{\zeta,\chi}\Gamma^{\rm{nl}} (k, \theta_{\alpha\beta})\Delta\sigma~, \label{45} 
\end{eqnarray}
where $\Delta\sigma$ is the size of grids in the parameter space $(\zeta,\chi)$. For example, we have $\Delta\sigma=0.018$ for the grids in the left panel of Fig.~\ref{F2}. Thus, the $\Gamma^{\rm{nl}} (k, \theta_{\alpha\beta})$ is proportional to the parameter $\kappa(k;\zeta,\chi)$ shown in Eq.~(\ref{43}). 
Here, we phenomenologically show the ORFs in Eq.~(\ref{45}) by letting $|\kappa|\equiv1$ on the left panel of Fig.~\ref{F8}. Because of the parameter space $\zeta,\chi\in (0,\infty)$ in Eq.~(\ref{32}), it is not practical to consider all the shapes of non-Gaussianity with $|\kappa|=1$. 
In principle, if the deviation from the Hellings-Downs curve is completely ascribed to the non-linear corrections of the ORFs, one can fit the $\kappa(k;\zeta,\chi)$ with real data \cite{NANOGrav:2020bcs}.

\section{conclusions and discussions} \label{V}
In this paper, we extended the study on the non-linear corrections of the ORFs in the present non-Gaussianity, in which self-interaction of gravity is first taken into consideration. 
Due to the self-interaction of gravity, the linear order GWs can generate the non-linear one, which consequently alters the response of GW detectors. Based on the perturbed Einstein field equations, and perturbed geodesic equations to the second order, we obtained non-linear order timing residuals of pulsar timing, and compute the ORFs with non-linear corrections in the PTA frequency band. 

 With parameterization of non-Gaussianity used in the present study, it is found that non-linear correction of ORFs might be dominated by the non-Gaussianity in the shape of the straight line, $\zeta + \chi =1$. All the contributions from the non-Gaussianity with large shape parameters $\zeta$ or $\chi$ are shown to be suppressed. Besides, the parameterization scenario might also result in the non-linear correlations of the ORFs being dependent on the sky position of pulsar pairs. It is analogous to the study on ORFs for anisotropic SIGWs \cite{Mingarelli:2013dsa}. Further discussions are elaborated upon in appendix~\ref{A}.

We considered the self-interaction of gravity by evaluating Einstein field equations in vacuum for the second-order metric perturbations. Namely, the space-time fluctuations are freely propagating within the GW detectors described in Einstein's gravity. It is suggested that the influence from the secondary effect of GWs on the detectors could be different in the alternative theory of gravity, or in the presence of (dark) matter. 

This paper showed that the leading order non-linear corrections for the ORFs come from the three-point correlations of $h_{{\bm k},ij}^{(1)}$. It is different from the pioneers' study that the correlations are from the four-point functions \cite{Tasinato:2022xyq}. It is because the contributions from three-point correlations in our study are all derived from the self-interaction of gravity shown in Eqs.~(\ref{21a})--(\ref{21e}), which was not considered in the pioneers' study.

\smallskip 
{\it Acknowledgments. } 
The author thanks Prof.~Qing-Guo Huang and Prof. Sai Wang for useful discussions.

\appendix

\appendix
\section{Dependence of location of pulsar pairs on the sky} \label{A}
\begin{figure}
	\includegraphics[width=\linewidth]{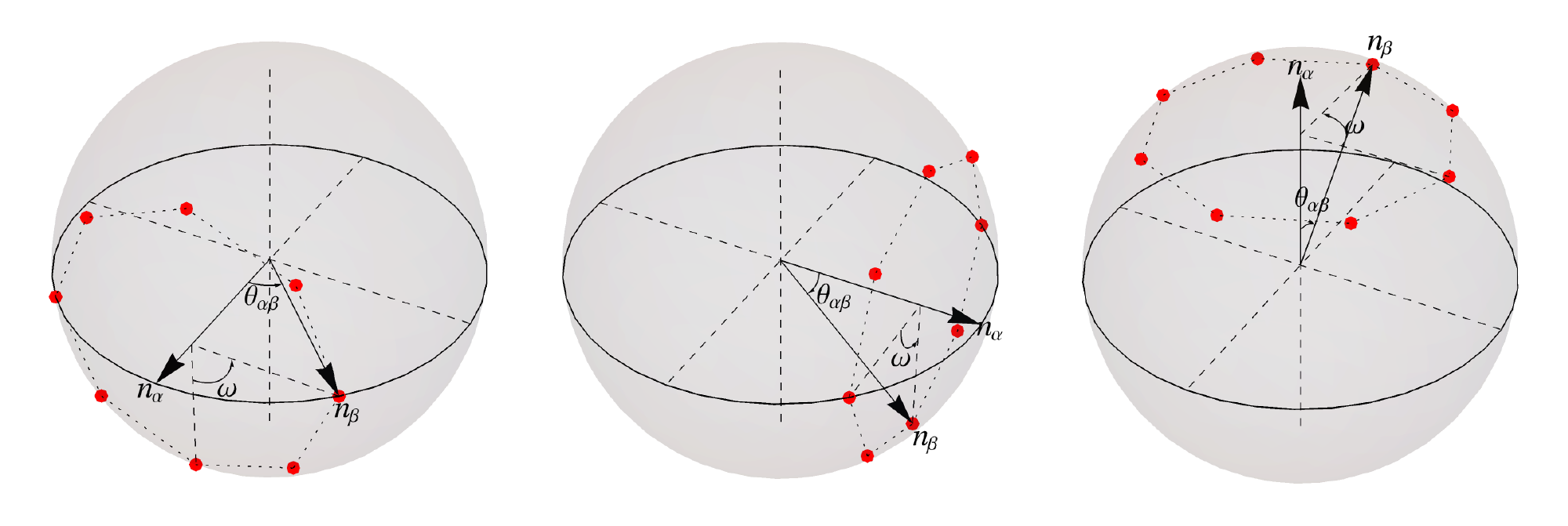}
	\caption{The schematic diagrams for selected $n_{\alpha}$ and $n_{\beta}$\label{F9}. The $n_\alpha$ are set to be $(1,0,0)$, $(0,1,0)$ and $(0,0,1)$, respectively. And the $\omega$ is the azimuthal angle of $n_\beta$ with respect to the $n_\alpha$, Different values of $\omega$ are represented by the points. }
\end{figure}

We computed non-linear correlations of ORFs as function the $\theta_{\alpha\beta}$ by letting $n_\alpha=(0,0,1)$ and $n_\beta=(\sin(\theta_{\alpha\beta}),0,\cos(\theta_{\alpha\beta}))$. It might disregard the possibility that the ORFs depend on both $n_\alpha$ and the $n_\beta$, rather than on $\theta_{\alpha\beta}$ alone. To clarify it, we here compute non-linear correlations of ORFs with different setups of $n_\alpha$ and $n_\beta$, as shown in the schematic diagrams in Fig.~\ref{F9}. 

\begin{figure}
	\includegraphics[width=\linewidth]{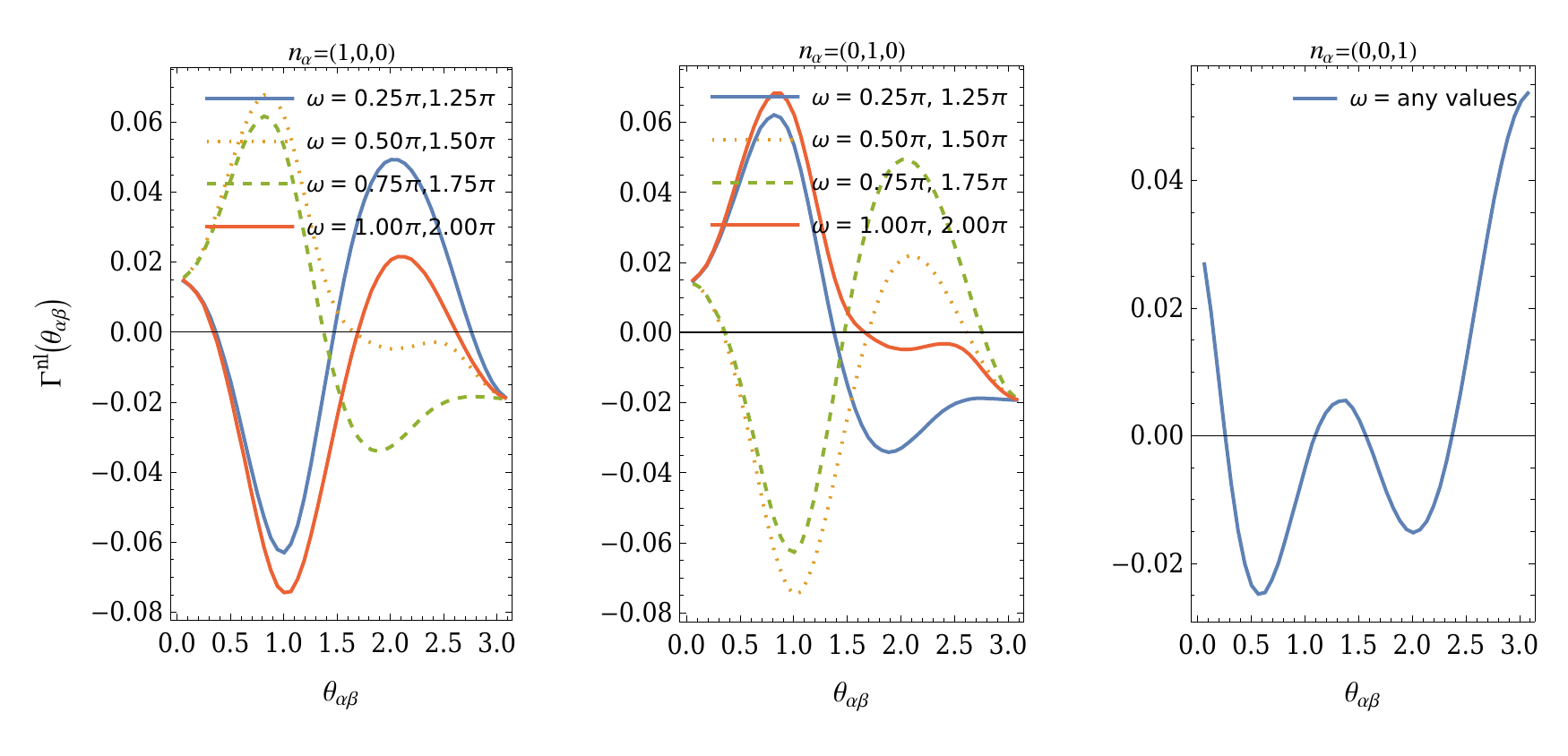}
	\caption{Non-linear correction of ORFs as function of $\theta_{\alpha\beta}$ with $\zeta+\chi=1$ for different setups of $n_\alpha$ and $n_\beta$. \label{F10}}
\end{figure}

For the shape of non-Gaussianity $\zeta=\chi=1$, the non-linear correlations of ORFs are presented in Fig.~\ref{F10}. It is found that the $\Gamma_{\text{nl}}$ depends on both $n_\alpha$ and $n_\beta$. To be specific, the $\Gamma_{\text{nl}}$ is shown to be axisymmetric with respect to $z$-axis. We further verify that the axisymmetry of ORFs even remains in the case of $\zeta\neq\chi$.
For the shape of non-Gaussianity $\zeta=\chi=1/2$, it is confirmed that the $\Gamma_{\text{nl}}$ is independent of $n_\alpha$ and azimuthal angle $\omega$ given in Fig.~\ref{F9}. This case is consistent with the pioneers' study \cite{Tasinato:2022xyq} that the non-linear correlations of ORFs only depend on the $\theta_{\alpha\beta}$.

\begin{figure}
	\includegraphics[width=\linewidth]{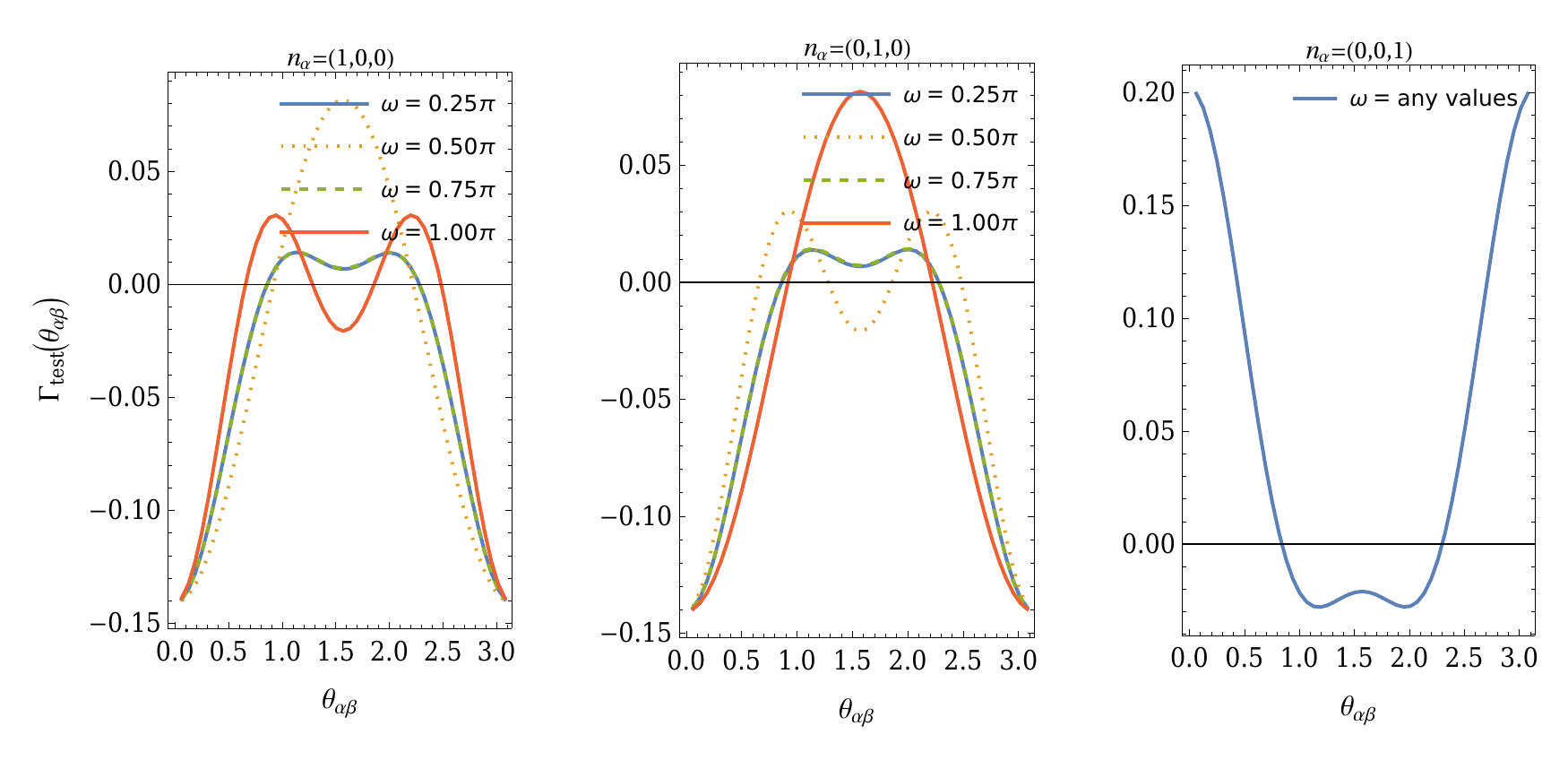}
	\caption{The $\Gamma_{\rm test}$ as function of $\theta_{\alpha\beta}$ for different setups of $n_\alpha$ and $n_\beta$.\label{F11}}
\end{figure}

It is noted that there is $n_\alpha$ and $n_\beta$-dependence of the non-linear correlations of ORFs, if polarization tensors $e_{ab}^\lambda(\hat{p})$, and $\hat{p}\neq\hat{k}$, are presented in angular integration in Eq.~(\ref{40}). This situation is analogous to the study on ORFs for anisotropic SIGWs \cite{Mingarelli:2013dsa}, in which there is an angular distribution of power spectrum in the angular integrations. To clarify this point, we further compute the integrations as follows,
\begin{eqnarray}
	\Gamma_{\rm test} (k, \theta_{\alpha \beta}) & = & \int \frac{{\rm d}
		\Omega}{4 \pi} \int \frac{{\rm d} \phi_q}{2 \pi} \left\{ \left(e^+_{m  l} (\hat{k}) e^+_{c  d}
	(\widehat{k - q}) e^+_{a  b} (\hat{q}) + e^{\times}_{m  l}
	(\hat{k}) e^{\times}_{c  d} (\widehat{k - q}) e^{\times}_{a
			b} (\hat{q})\right)
		\right. \nonumber\\ && \left. \times
	\left( \mathcal{K}_{\rm test}^{0, m  l} \left( \bm{k}, \hat{n}_{\alpha}
		\right) \mathcal{F}_{\rm test}^{c  d, a  b} \left( \bm{k},
		\bm{q}, \hat{n}_{\beta} \right)
	+\mathcal{F}_{\rm test}^{c  d, a b} \left( \bm{k}, \bm{q}, \hat{n}_{\alpha} \right) \mathcal{K}_{\rm test}^{0, m l} \left( \bm{k}, \hat{n}_{\beta} \right) \right) \right\}~. 
\end{eqnarray}
where $\mathcal{K}_{\rm test}^{0, a  b}\left( \bm{k}, \hat{n}
		\right)=\hat{n}^a\hat{n}^b$ and $\mathcal{F}_{\rm test}^{c  d, a b} \left( \bm{k}, \bm{q}, \hat{n} \right)=\hat{n}^c\hat{n}^d\hat{n}^a\hat{n}^b$. It is found that the $\Gamma_{\rm test}$ as function of $\theta_{\alpha\beta}$ also depends on both $n_\alpha$ and $n_\beta$, and shown in Fig.~\ref{F11}. It suggests that $n_\alpha$ and $n_\beta$-dependence of the non-linear correlations of ORFs is the intrinsic prosperity of the polarization tensors, and thus it is inevitable. 

\section{Non-Gaussianity of scalar induced GWs} \label{B}
The non-Gaussianity could be produced from the non-linear generation mechanism of GWs in cosmology, even when the source of GWs itself is Gaussian. Here, we will provide an example. It is known that
the small-scale primordial curvature perturbations that re-enter the horizon after the inflation epoch might lead to the generation of secondary GWs \cite{Baumann:2007zm,Wang:2016ana,Kohri:2018awv,Bartolo:2018evs,Sasaki:2018dmp,Chen:2019xse}. The equation of the GWs, $h_{ij}$, is presented as follows,
\begin{eqnarray}
  h''_{i  j} + 2\mathcal{H}h'_{i  j} - \Delta h_{i  j}
  & = & - 4 \Lambda_{i  j}^{a  b} \left( 3 \partial_a \phi
  \partial_b \phi + \frac{1}{\mathcal{H}} (\partial_a \phi' \partial_b \phi +
  \partial_a \phi \partial_b \phi') + \frac{1}{\mathcal{H}^2} \partial_a \phi'
  \partial_b \phi' \right)~,
\end{eqnarray}
where the scalar perturbations $\phi$ are proportional to the primordial curvature perturbations, with a constant coefficient.
The equations of $h_{i  j, \bm{k}}$, which is the Fourier
mode of $h_{i  j}$, also can be obtained,
\begin{eqnarray}
  h_{i  j, \bm{k}}'' + 2\mathcal{H}  h_{i  j,
  \bm{k}}' + k^2 h_{i  j, \bm{k}} & = & 4 \Lambda^{a 
  b}_{i  j} \mathcal{S}_{a  b, \bm{k}}~, \label{B2}
\end{eqnarray}
where the Fourier mode of $\phi$ is $\phi_{\bm{k}} \left( =
\Phi_{\bm{k}} T (k \eta) \right)$ and,
\begin{eqnarray}
  \mathcal{S}_{a  b, \bm{k}} & = & \int \frac{{\rm d}^3 p}{(2
  \pi)^3} \bigg\{ (k_a - p_a) p_b \Phi_{\bm{k} - \bm{p}}
  \Phi_{\bm{p}} \bigg( 3 T \left( \left| \bm{k} - \bm{p} \right| \eta \right)
  T (p \eta) \nonumber\\
  & & \left. \frac{T' \left( \left| \bm{k} - \bm{p} \right| \eta
  \right) T (p \eta) + T \left( \left| \bm{k} - \bm{p} \right| \eta
  \right) T' (p \eta)}{\mathcal{H}} + \frac{T' \left( \left| \bm{k} -
  \bm{p} \right| \eta \right) T' (p \eta)}{\mathcal{H}^2} \right) \bigg\}~.
\end{eqnarray}
Solving Eq.~(\ref{B2}), the $h_{ij,\bm k}$ can be written in the form of
\begin{eqnarray}
  h_{i  j, \bm{k}} & = & \int \frac{{\rm d}^3 p}{(2 \pi)^3} \left\{
  \Lambda^{a  b}_{i  j} \left( \bm{k} \right) p_a p_b I
  \left( \left| \bm{k} - \bm{p} \right|, p, \eta \right)
  \Phi_{\bm{k} - \bm{p}} \Phi_{\bm{p}} \right\}~,\label{B4}
\end{eqnarray}
where
\begin{eqnarray}
  I \left( \left| \bm{k} - \bm{p} \right|, p, \eta \right) & = & 4
  \int_0^{\eta} {\rm d} \bar{\eta}  G_k (\eta, \bar{\eta}) f \left(
  \left| \bm{k} - \bm{p} \right|, p, \eta \right)~.
\end{eqnarray}
For illustration, we let $h_{1,\bm k}\equiv h_{ij,\bm k}$ in the following.
Under the assumptions that the primordial curvature perturbations $\Phi_{\bm k}$ are isotropic, unpolarized, and Gaussian, the two-point functions of $\Phi_{\bm k}$ can take the form of 
\begin{eqnarray}
  \left\langle \Phi_{\bm{k}} \Phi_{\bm{k}'} \right\rangle & = & (2
  \pi)^3 \delta \left( \bm{k} + \bm{k}' \right) P_{\Phi} (k)~.\label{B6}
\end{eqnarray}
Therefore, based on Eqs.~(\ref{B4}) and (\ref{B6}), the two-point function of $h_{i  j, \bm{k}}$ is shown to be
\begin{eqnarray}
  \left\langle h_{1, \bm{k}_1} h_{2, \bm{k}_2} \right\rangle & = & (2
  \pi)^3 \delta \left( \bm{k}_1 + \bm{k}_2 \right) P_{h, 12} \left(
  \bm{k}_1, \eta \right)~,
\end{eqnarray}
where 
\begin{eqnarray}
  P_{h, 12} (\bm{k}_1, \eta ) & = & \Lambda^{a  b}_1
  ( \bm{k}_1 ) \Lambda^{c  d}_2 (- \bm{k}_1)
   \int \frac{{\rm d}^3 p}{(2 \pi)^3} \big\{ P_{\Phi} \left( \left|
  \bm{k}_1 - \bm{p} \right| \right) P_{\Phi} (p) p_a p_b p_c p_d \nonumber\\
  & & \times \left( I (\left| \bm{k}_1 - \bm{p}
  \right|, p, \eta) I (\left| \bm{k}_1 - \bm{p} \right|,
  p, \eta) + I (\left| \bm{k}_1 - \bm{p} \right|, p, \eta
  )I ( p, \left| \bm{k}_1 - \bm{p} \right|, \eta ) 
  \right) \big\}~.
\end{eqnarray} 
And the there-point function of $h_{i  j, \bm{k}} $ is
\begin{eqnarray}
  \left\langle h_{1, \bm{k}_1} h_{2, \bm{k}_2} h_{3, \bm{k}_3}
  \right\rangle & = & (2 \pi)^3 \delta \left( \bm{k}_1 + \bm{k}_2 +
  \bm{k}_3 \right) \mathcal{B}_{h, 123} \left( \bm{k}_1, \bm{k}_2,
  \eta \right)~, 
\end{eqnarray}
where 
\begin{eqnarray}
  \mathcal{B}_{h, 123} \left( \bm{k}_1, \bm{k}_2, \eta \right) & = &
  \Lambda^{a  b}_1 \left( \bm{k}_1 \right) \Lambda^{c 
  d}_2 \left( \bm{k}_2 \right) \Lambda^{m  n}_3 \left(
  \bm{k}_3 \right) \int \frac{{\rm d}^3 p}{(2 \pi)^3} \bigg\{ p_a p_b (k_{1, c} - p_c) (k_{1, d} - p_d) p_m p_n \nonumber\\
  && \times P_{\Phi} \left(
  \left| \bm{k}_1 - \bm{p} \right| \right) P_{\Phi} (p) P_{\Phi}
  \left( \left| \bm{k}_1 + \bm{k}_2 - \bm{p} \right| \right) I
  \left( \left| \bm{k}_1 - \bm{p} \right|, p, \eta \right) \nonumber\\
  && \times \left( I
  \left( \left| \bm{k}_1 + \bm{k}_2 - \bm{p} \right|, \left|
  \bm{k}_1 - \bm{p} \right|, \eta \right) + I \left( \left|
  \bm{k}_1 - \bm{p} \right|, \left| \bm{k}_1 + \bm{k}_2 -
  \bm{p} \right|, \eta \right) \right) \nonumber\\
  && \times \left( I \left( \left| 2
  \bm{k}_1 + \bm{k}_2 - \bm{p} \right|, \left| \bm{k}_1 +
  \bm{k}_2 - \bm{p} \right|, \eta \right) + I \left( \left|
  \bm{k}_1 + \bm{p} \right|, p \right) \right) \nonumber\\
  & & + p_a p_b p_c p_d (k_{1, m} - p_m) (k_{1, n} - p_n) P_\Phi \left( \left|
  \bm{k}_1 - \bm{p} \right| \right) P_\Phi (p) P_\Phi \left( \left| \bm{k}_2
  + \bm{p} \right| \right) \nonumber\\
  && \times I \left( \left| \bm{k}_1 - \bm{p}
  \right|, p, \eta \right) \left( I \left( p, \left| \bm{k}_2 + \bm{p}
  \right|, \eta \right) + I \left( \left| \bm{k}_2 + \bm{p} \right|,
  p, \eta \right) \right) \nonumber\\
  && \times \left( I \left( \left| 2 \bm{k}_1 - \bm{p}
  \right|, \left| \bm{k}_1 - \bm{p} \right|, \eta \right) + I \left(
  \left| \bm{k}_1 + \bm{k}_2 + \bm{p} \right|, \left| \bm{k}_2
  + \bm{p} \right|, \eta \right) \right) \bigg\}~. \label{B10}
\end{eqnarray}
It indicated that the $h_{ij,\bm k}$ is a non-Gaussian variable, because the bispectrum does not vanish. 

\bibliography{ref}
\end{document}